\documentclass[10pt,twocolumn,letterpaper]{article}

\usepackage{cvpr}              

\usepackage[dvipsnames]{xcolor}

\usepackage{colortbl}
\usepackage{multirow}
\usepackage{makecell}
\usepackage{booktabs}
\usepackage{pifont}
\usepackage{hhline}
\usepackage{algorithm}
\usepackage{algorithmic}

\usepackage{soul}

\definecolor{cvprblue}{rgb}{0.21,0.49,0.74}
\usepackage[pagebackref,breaklinks,colorlinks,citecolor=cvprblue]{hyperref}

\def\paperID{12911} 
\def\confName{CVPR}
\def\confYear{2024}

\title{DisDet: Exploring Detectability of Backdoor Attack on Diffusion Models}

\author{ Yang Sui\textsuperscript{1} \quad Huy Phan\textsuperscript{1} \quad Jinqi Xiao\textsuperscript{1} \quad Tianfang Zhang\textsuperscript{1} \quad Zijie Tang\textsuperscript{2} \quad \\
Cong Shi\textsuperscript{3} \quad Yan Wang\textsuperscript{2} \quad Yingying Chen\textsuperscript{1} \quad Bo Yuan\textsuperscript{1} \\
\textsuperscript{1}Rutgers University \quad \textsuperscript{2}Temple University \quad \textsuperscript{3}New Jersey Institute of Technology
}

\begin{document}
\maketitle
\begin{abstract}

In the exciting generative AI era, the diffusion model has emerged as a very powerful and widely adopted content generation and editing tool for various data modalities, making the study of their potential security risks very necessary and critical. Very recently, some pioneering works have shown the vulnerability of the diffusion model against backdoor attacks, calling for in-depth analysis and investigation of the security challenges of this popular and fundamental AI technique.

In this paper, for the first time, we systematically explore the detectability of the poisoned noise input for the backdoored diffusion models, an important performance metric yet little explored in the existing works. Starting from the perspective of a defender, we first analyze the properties of the trigger pattern in the existing diffusion backdoor attacks, discovering the important role of distribution discrepancy in Trojan detection. Based on this finding, we propose a low-cost trigger detection mechanism that can effectively identify the poisoned input noise. We then take a further step to study the same problem from the attack side, proposing a backdoor attack strategy that can learn the unnoticeable trigger to evade our proposed detection scheme. 

Empirical evaluations across various diffusion models and datasets demonstrate the effectiveness of the proposed trigger detection and detection-evading attack strategy. For trigger detection, our distribution discrepancy-based solution can achieve a 100\% detection rate for the Trojan triggers used in the existing works. For evading trigger detection, our proposed stealthy trigger design approach performs end-to-end learning to make the distribution of poisoned noise input approach that of benign noise, enabling nearly 100\% detection pass rate with very high attack and benign performance for the backdoored diffusion models. 


\end{abstract}    
\section{Introduction}
\label{sec:intro}

Recently, the diffusion model has emerged as a prevalent generative AI technique for content creation and editing across various data modalities, including image, video, speech, text, etc. Built on the core principle originating from non-equilibrium thermodynamics, a diffusion model aims to learn to generate the target probability distribution via constructing and reverting a series of latent variables. Thanks to its solid theoretical foundations and training stability, to date the diffusion models have been widely used in various generative tasks, such as image generation \cite{ho2020denoising, song2020denoising, bordes2022high, chao2021denoising, karras2022elucidating}, text-to-image synthesis \cite{rombach2022high, saharia2022photorealistic, ramesh2022hierarchical, singh2023high}, image editing \cite{meng2021sdedit, couairon2022diffedit, brooks2023instructpix2pix}, image inpainting \cite{lugmayr2022repaint}, super-resolution \cite{rombach2022high, saharia2022image, choi2021ilvr} and video generation \cite{ho2022imagen, esser2023structure, harvey2022flexible}.

Because diffusion models have already served as the backbone components in many real-world applications, the corresponding security issues have become a potentially challenging risk that requires special attention. In particular, the vulnerabilities of diffusion models under \textit{backdoor attack}, as a common and essential attack strategy against the existing classification models \cite{chen2017targeted, gu2019badnets, nguyen2020wanet, liu2020reflection, doan2021lira, zheng2023trojvit, yuan2023you}, should be carefully reviewed and studied in the emerging generative AI era. 

Despite the current research prosperity of the applications of diffusion models, the security challenges of this vital technique in the backdoor attack scenario are still under-explored. To date, only very few works investigate the backdoor attack tailored to diffusion models. In particular, \cite{chen2023trojdiff,chou2023backdoor}, as the representative works in this topic, propose the forward and backward processes of the backdoored diffusion models, demonstrating that the currently representative diffusion models can be attacked to generate the images in a target category or even targeted fixed images, with the presence of poisoned input noise.

Although these existing works reveal the feasibility of implanting Trojans into the diffusion models, we argue that the study on the vulnerability and robustness of diffusion models under backdoor attack is still in its infancy. In particular, the prior efforts use the fixed trigger pattern to craft the poisoned noise input, leaving a blank in exploring the \textit{detectability} of the embedded Trojans, a critical metric directly affecting the overall attack performance. From the perspective of AI security research, such exploration of the stealthiness of the Trojan trigger is very critical in both attacker and defender aspects.

Aiming to fill this research blank and obtain a deep understanding of the behavior of diffusion models against backdoor attacks, in this paper, we propose a systematic study on the detectability of Trojan input for the backdoored diffusion models, from both attacker and defender perspectives. We first analyze the characteristics of the existing fixed trigger pattern and discover that the distribution discrepancy of noise input can be used as a good marker. Based on this finding, we develop a low-cost trigger detection mechanism that can effectively identify the poisoned input noise. We then take a further step to propose the backdoor attack strategy that can learn the stealthy trigger evading the proposed detection scheme, enriching the research on the security of diffusion models. Overall, the contributions of this paper are summarized as follows:

\begin{itemize}
    \item We explore the detectability of trigger patterns in the state-of-the-art diffusion model backdoor attacks. By analyzing the distribution discrepancy of the noise input, we propose a distribution-based detection mechanism that can identify the poisoned noise input of the backdoored diffusion models at a low cost.
    \item We then develop a backdoor attack strategy that can evade our proposed detection method. By performing end-to-end learning of the trigger pattern towards minimizing the distribution discrepancy, the poisoned noise input can exhibit a very similar distribution to the benign input, making the backdoor attack unnoticeable. We also optimize the training process of the stealthy trigger pattern to improve the benign and attack performance of the backdoored diffusion models.
    \item We perform empirical evaluations for different diffusion models across different datasets and demonstrate the effectiveness of the proposed trigger detection and detection-evading attack strategy. On the defender side, our proposed distribution-based detection method can achieve a 100\% detection rate for the trigger patterns used in the existing works. On the attacker side, our proposed detection-evading trigger can enable nearly 100\% detection pass rate and bring high attack and benign performance for the backdoored diffusion models. 
\end{itemize}

\section{Related Works}
\label{sec:relatedworks}

\noindent \textbf{Diffusion Models.} Diffusion models have emerged as a powerful generative AI technique very recently. Compared with other deep generative models, diffusion models exhibit good training stability and better quality and diversity of the generated data, making them popularly adopted in a variety of generative tasks, e.g., image generation \cite{ho2020denoising, song2020denoising, ho2022cascaded, dhariwal2021diffusion, liu2023more, bordes2022high, chao2021denoising, karras2022elucidating}, video generation \cite{ho2022imagen, esser2023structure, harvey2022flexible}, text-to-image synthesis \cite{rombach2022high, saharia2022photorealistic, ramesh2022hierarchical, singh2023high, kumari2023multi, gu2022vector, zhang2023adding, ruiz2023dreambooth, zhang2023adding} and fast sampling \cite{song2020denoising, salimans2021progressive, lu2022dpm}. Diffusion models can be formulated in different ways, such as denoising diffusion probabilistic model (DDPM) \cite{ho2020denoising} and its variant DDIM \cite{song2020denoising}, noise conditional score network (NCSN)  \cite{song2020score} and latent diffusion model (LDM) \cite{rombach2022high}. This paper focuses on the backdoor attack on DDPM/DDIM, as the most representative and fundamental diffusion model type.


\noindent \textbf{Backdoor Attacks on AI Models.} The research on launching backdoor attacks against AI models, especially the classification models, has been widely reported in the literature \cite{chen2017targeted, gu2019badnets, nguyen2020wanet, liu2020reflection, doan2021lira, saha2020hidden, li2021invisible, salem2022dynamic, nguyen2020input}. In this attack scenario, the adversary first poisons the training data to inject the backdoor into the model in the training phase. Then in the inference phase, the backdoored model behaves normally with the presence of benign input; while it will exhibit malicious behavior (e.g., misclassification) when the input is embedded with a Trojan trigger. Considering its natural stealthiness and severe damage, a series of backdoor defense approaches have been proposed \cite{gao2019strip, wang2019neural, liu2018fine, chen2019detecting, chen2019deepinspect, tran2018spectral, li2020rethinking, doan2023defending}.

\noindent \textbf{Backdoor Attacks on Diffusion Models.} Unlike the extensive research on classification models, the backdoor attack for diffusion models is little explored yet. To date, the most two representative works are \cite{chen2023trojdiff,chou2023backdoor}, which for the first time demonstrate the feasibility of launching backdoor attack against the generative models. By adding a pre-defined trigger into the benign Gaussian noise input, the manipulated poisoned noise can prompt the backdoored diffusion model to generate a target image \cite{chou2023backdoor, chen2023trojdiff} (e.g., Hello Kitty)  or images belonging to a certain class \cite{chen2023trojdiff} (e.g., ``horse")  as desired by attackers. Because the adversary can leverage such malicious behavior to generate potentially offensive or illegal images, the vulnerability of diffusion models against backdoor attacks poses severe security challenges and risks.



\section{Background}
\label{sec:background}

\subsection{Diffusion Model}
\label{subsec:dm}

Diffusion model \cite{sohl2015deep, ho2020denoising} is a type of deep generative model aiming to generate semantic-rich data from Gaussian noise. To realize such mapping, a diffusion model typically consists of forward \textit{diffusion process} and backward \textit{generative process}. Take the representative denoising diffusion probabilistic model (DDPM) \cite{ho2020denoising} as an example. In the diffusion process, an image $\mathbf{x}_0$ sampled from real data distribution $q(\mathbf{x}_0)$ is gradually diffused with the added random Gaussian noise over $T$ time steps. More specifically, this procedure generates a sequence of random variables $\mathbf{x}_1, \mathbf{x}_2, \cdots, \mathbf{x}_T$ in a Markov chain as $\mathbf{x}_t=\sqrt{1-\beta_t}\mathbf{x}_{t-1}+\beta_t\boldsymbol{\epsilon}$ and $q(\mathbf{x}_t|\mathbf{x}_{t-1}) := \mathcal{N}(\mathbf{x}_t; \sqrt{1-\beta_t}\mathbf{x}_{t-1},\beta_t\mathbf{I})$, where $\beta_t$ is the pre-defined variance schedule and $\boldsymbol{\epsilon} \sim \mathcal{N}(0, \mathbf{I})$. For simplicity, by defining $\alpha_t = 1 - \beta_t$ and $\overline{\alpha}_t = \prod \nolimits_{i=1}^t \alpha_i$, the diffusion process can be formulated as $q(\mathbf{x}_t | \mathbf{x}_0) = \mathcal{N}(\mathbf{x}_t; \sqrt{\overline{\alpha}_t}\mathbf{x}_{0},(1-\overline{\alpha}_t)\mathbf{I})$. Then in the generative process, a parameterized Markov chain is trained aiming to
reverse the diffusion process and recover the image from the noise. To be specific, it learns model parameters $\theta$ such that the reverse transition $p_\theta(\mathbf{x}_{t-1} | \mathbf{x}_t)$, which is defined as $\mathcal{N}(\mathbf{x}_{t-1}; \boldsymbol{\mu}_\theta(\mathbf{x}_t, t), \boldsymbol{\Sigma}_\theta(\mathbf{x}_t, t))$, is equivalent to the forward transition $q(\mathbf{x}_{t-1} | \mathbf{x}_t, \mathbf{x}_0)=\mathcal{N}(\mathbf{x}_{t-1}; \hat{\boldsymbol{\mu}}_t(\mathbf{x}_t, \mathbf{x}_0), \hat{\beta}_t \mathbf{I})$, where $\hat{\boldsymbol{\mu}}_t(\mathbf{x}_t, \mathbf{x}_0) = \frac{1}{\sqrt{\alpha_t}}((\sqrt{\overline{\alpha}_t}\mathbf{x}_{0} + (1-\overline{\alpha}_t)\boldsymbol{\epsilon})-\frac{\beta_t}{\sqrt{1 - \overline{\alpha}_t}}\boldsymbol{\epsilon})$. To that end, DDPM aims to align the mean between $p_\theta(\mathbf{x}_{t-1} | \mathbf{x}_t)$ and $q(\mathbf{x}_{t-1} | \mathbf{x}_t, \mathbf{x}_0)$ via minimizing the following training objective:
\begin{equation}
\begin{aligned}
\mathbb{E}_{t, \mathbf{x}_0, \boldsymbol{\epsilon}} [\left\| \boldsymbol{\epsilon} - \boldsymbol{\epsilon}_{\theta}(\sqrt{\overline{\alpha}_{t}}\mathbf{x}_0 + \sqrt{1-\overline{\alpha}_{t}}\boldsymbol{\epsilon}, t) \right\|^{2}], \\
\end{aligned}
\label{eqn:dm_objective}
\end{equation}
where $t$ is uniformly sampled from  ${\{ 1,\cdots,T\}}$. Here, $\boldsymbol{\epsilon}_{\theta}$ represents a parameterized denoiser to predict noise $\boldsymbol{\epsilon}$ from $\mathbf{x}_t$, which is usually implemented based on U-Net \cite{ronneberger2015u}. 


\subsection{Backdoor Attack on Diffusion Model}
\label{subsec:backdoorDM}

\noindent \textbf{Threat Model.} Following the settings in \cite{chen2023trojdiff, chou2023backdoor}, we assume that the attacker aims to train a backdoored diffusion model, which will 1) generate the clean image from the distribution $q(\mathbf{x}_0)$ with benign Gaussian noise input $\mathcal{N}(0, \mathbf{I})$; and 2) generate the target image from the distribution $\Tilde{\mathbf{x}}_0 \sim \Tilde{q}(\Tilde{\mathbf{x}}_0)$ with the presence of poisoned noise input $\tilde{\mathbf{x}}_T$ that is embedded with a pre-defined \textit{trigger} $\boldsymbol{\delta}$. Without loss of generality\footnote{We do not specifically study patch-based trigger because 1) patching can be viewed as special case of blending \cite{chen2023trojdiff}; and 2) patch-based trigger is more perceptible to visual inspection, making it less stealthy than blending-based trigger.}, we assume the trigger is proportionally blended to the clean Gaussian noise with propositional factor $\gamma \in [0, 1]$. More specifically, $\tilde{\mathbf{x}}_T \sim \mathcal{N}(\boldsymbol{\mu}_{\boldsymbol{\delta}}, \gamma^2\mathbf{I})$, where $\boldsymbol{\mu}_{\boldsymbol{\delta}} = (1-\gamma)\boldsymbol{\delta}$ satisfying $\tilde{\mathbf{x}}_T = (1-\gamma) \boldsymbol{\delta} + \gamma \boldsymbol{\epsilon}, \boldsymbol{\epsilon} \sim \mathcal{N}(0, \mathbf{I})$. 



\noindent \textbf{Backdoored Diffusion and Generative Processes.} To realize the attack goal, we assume that the adversary is allowed to modify the diffusion and generative processes and the training procedure. More specifically, as indicated in \cite{chen2023trojdiff}, the attacker first diffuses the distribution $\Tilde{q}(\Tilde{\mathbf{x}}_0)$ of the target images to $\mathcal{N}(\boldsymbol{\mu}_{\boldsymbol{\delta}}, \gamma^{2}\mathbf{I})$, forming a backdoored diffusion process as 
$\Tilde{q}(\Tilde{\mathbf{x}}_{t-1} | \Tilde{\mathbf{x}}_t, \Tilde{\mathbf{x}}_0)=\mathcal{N}(\Tilde{\mathbf{x}}_{t-1}; \Tilde{\boldsymbol{\mu}}_{t}(\Tilde{\mathbf{x}}_t, \Tilde{\mathbf{x}}_0), \Tilde{\beta}_{t} \mathbf{I})$, 
where $\Tilde{\boldsymbol{\mu}}_{t}(\Tilde{\mathbf{x}}_t, \Tilde{\mathbf{x}}_0) = \frac{1}{\sqrt{\alpha_t}}((\sqrt{\overline{\alpha}_t}\Tilde{\mathbf{x}}_{0} + \sqrt{1-\overline{\alpha}_t}\boldsymbol{\mu}_{\boldsymbol{\delta}} + \sqrt{1-\overline{\alpha}_{t}}\gamma\boldsymbol{\epsilon})-\frac{\beta_t}{\sqrt{1 - \overline{\alpha}_t}}\boldsymbol{\epsilon})$. Then in the generative process, the parameterized model $\theta$ is learned to reverse both the benign and backdoored diffusion processes: $p_\theta(\mathbf{x}_{t-1} | \mathbf{x}_t)=q(\mathbf{x}_{t-1} | \mathbf{x}_t, \mathbf{x}_0)$ (for benign Gaussian input case described in \cref{subsec:dm}) and $\Tilde{p}_\theta(\Tilde{\mathbf{x}}_{t-1} | \Tilde{\mathbf{x}}_t) = \mathcal{N}(\Tilde{\mathbf{x}}_{t-1}; \Tilde{\boldsymbol{\mu}}_\theta(\Tilde{\mathbf{x}}_t, t), \Tilde{\boldsymbol{\Sigma}}_\theta(\Tilde{\mathbf{x}}_t, t))=\Tilde{q}(\Tilde{\mathbf{x}}_{t-1} | \Tilde{\mathbf{x}}_t, \Tilde{\mathbf{x}}_0)$  (for poisoned noise input case). To that end, the corresponding training objectives aim to simultaneously optimize both the benign and backdoor diffusion processes. Specifically, the benign training objective follows \cref{eqn:dm_objective}, and the backdoored diffusion training objective is formulated as:
\begin{equation}
\begin{aligned}
\mathbb{E}_{t, \Tilde{\mathbf{x}}_0, \boldsymbol{\epsilon}} [\left\| \boldsymbol{\epsilon} - \boldsymbol{\epsilon}_{\theta}(\sqrt{\overline{\alpha}_{t}}\Tilde{\mathbf{x}}_0 + \sqrt{1-\overline{\alpha}_t}\boldsymbol{\mu}_{\boldsymbol{\delta}} + \sqrt{1-\overline{\alpha}_{t}}\gamma\boldsymbol{\epsilon}, t) \right\|^{2}], \\
\end{aligned}
\label{eqn:backdoor_dm_objective}
\end{equation}
where $\Tilde{\mathbf{x}}_0 \sim \Tilde{q}(\Tilde{\mathbf{x}}_0)$. Here $t, \boldsymbol{\epsilon}, \boldsymbol{\epsilon}_{\theta}$ are with the same setting in \cref{eqn:dm_objective}. 


\section{Trigger Detection in Backdoored Diffusion}
\label{sec:detection}

\begin{figure}[t]
  \centering
  \begin{subfigure}{0.32\linewidth}
    \includegraphics[width=1\linewidth]{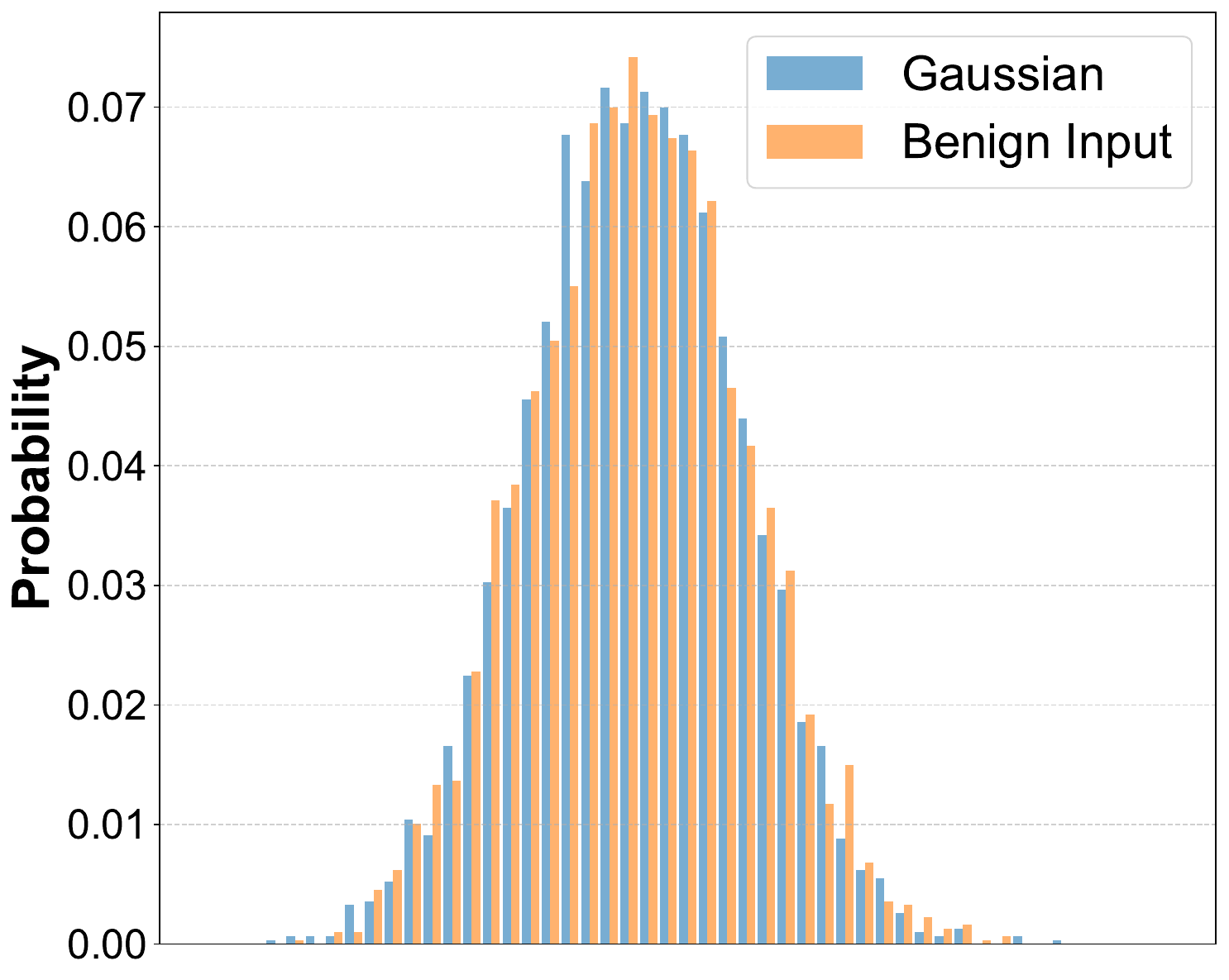}
    \caption{PDD score: 0.029.}
    \label{fig:distribution_gaussian_a}
  \end{subfigure}
  \begin{subfigure}{0.32\linewidth}
    \includegraphics[width=1\linewidth]{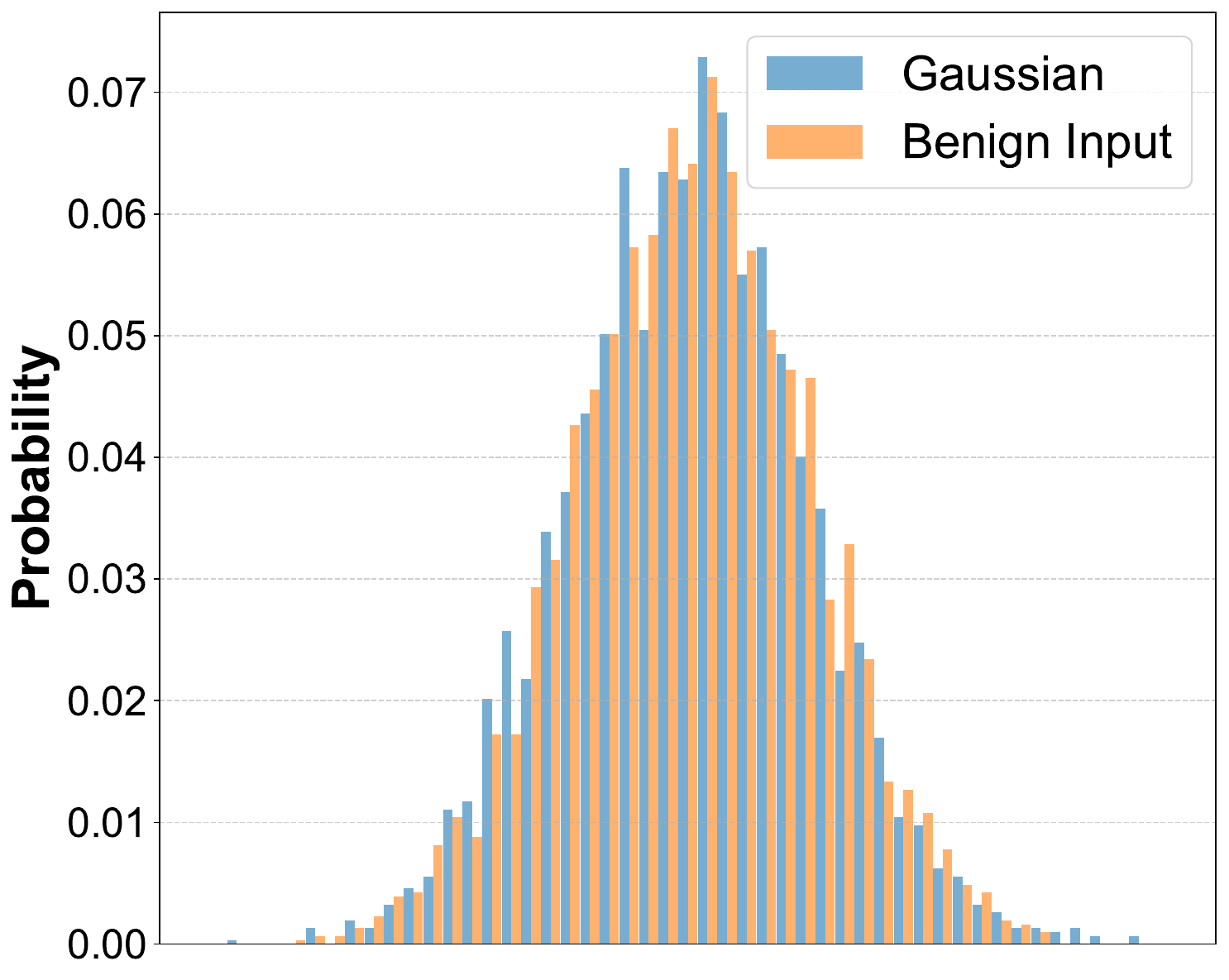}
    \caption{PDD score: 0.032.}
    \label{fig:distribution_gaussian_b}
  \end{subfigure}
    \begin{subfigure}{0.32\linewidth}
    \includegraphics[width=1\linewidth]{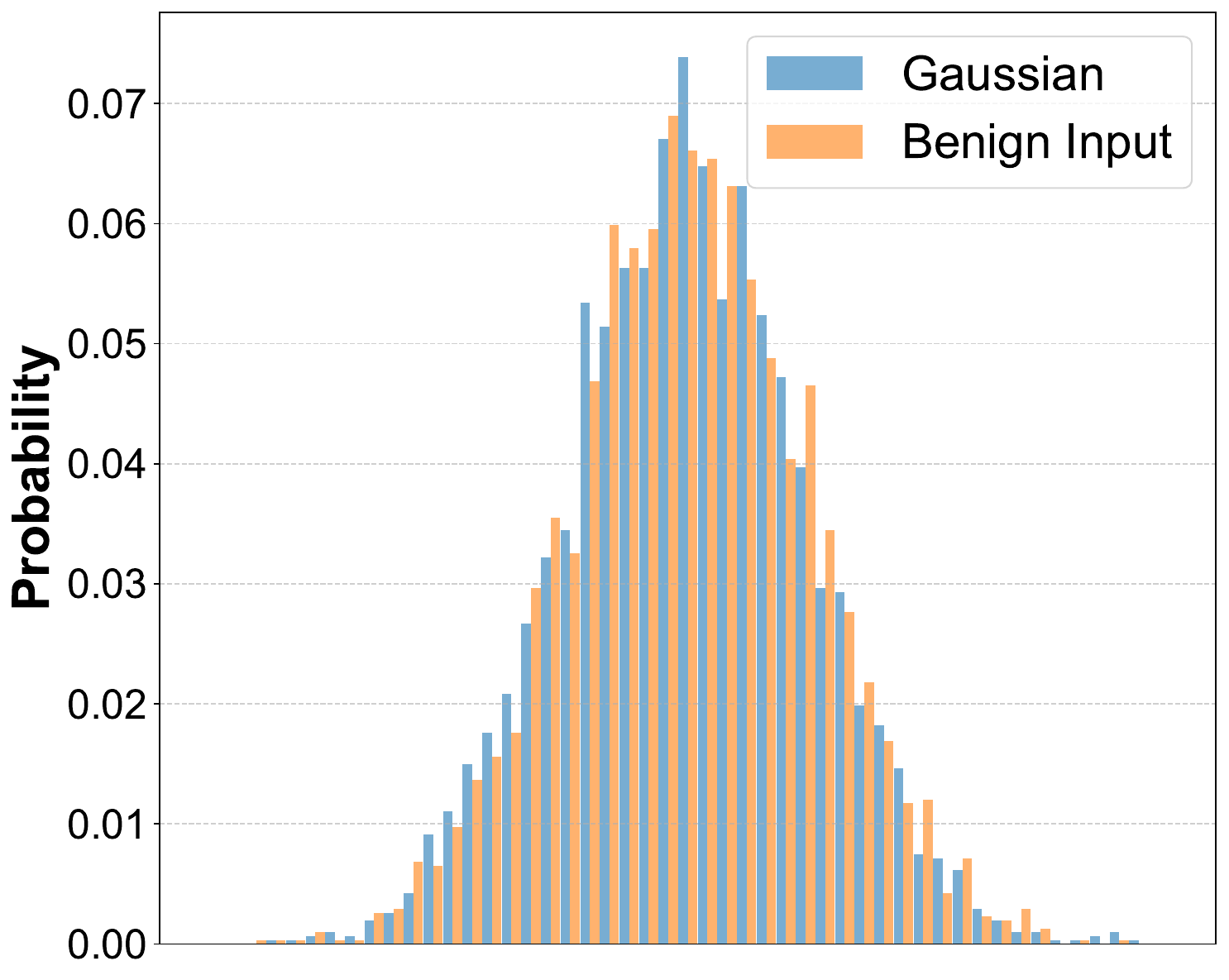}
    \caption{PDD score: 0.031.}
    \label{fig:distribution_gaussian_c}
  \end{subfigure}

    \vfill
    
      \begin{subfigure}{0.32\linewidth}
    \includegraphics[width=1\linewidth]{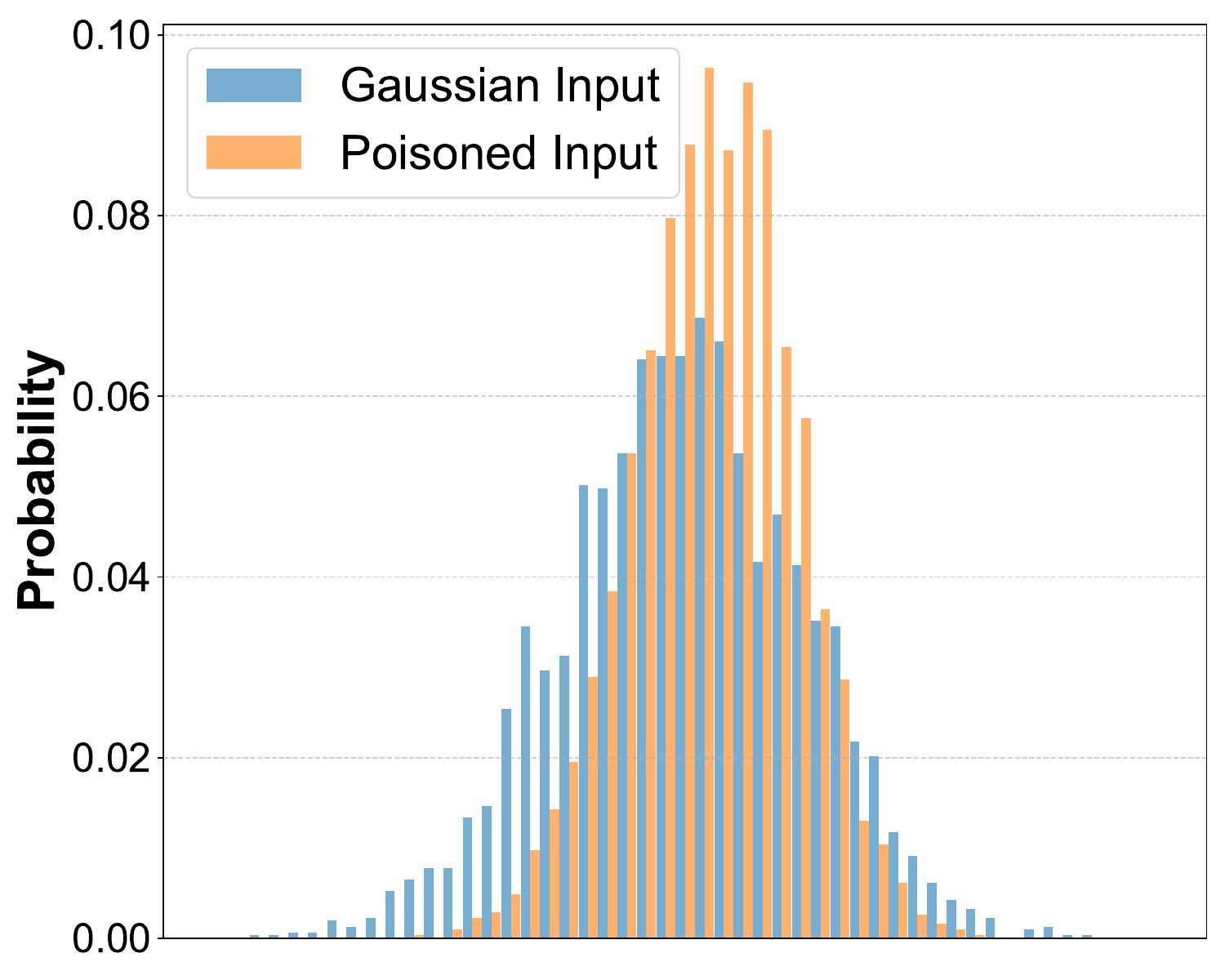}
    \caption{PDD score: 0.182.}
    \label{fig:distribution_blend_a}
  \end{subfigure}
  \begin{subfigure}{0.32\linewidth}
    \includegraphics[width=1\linewidth]{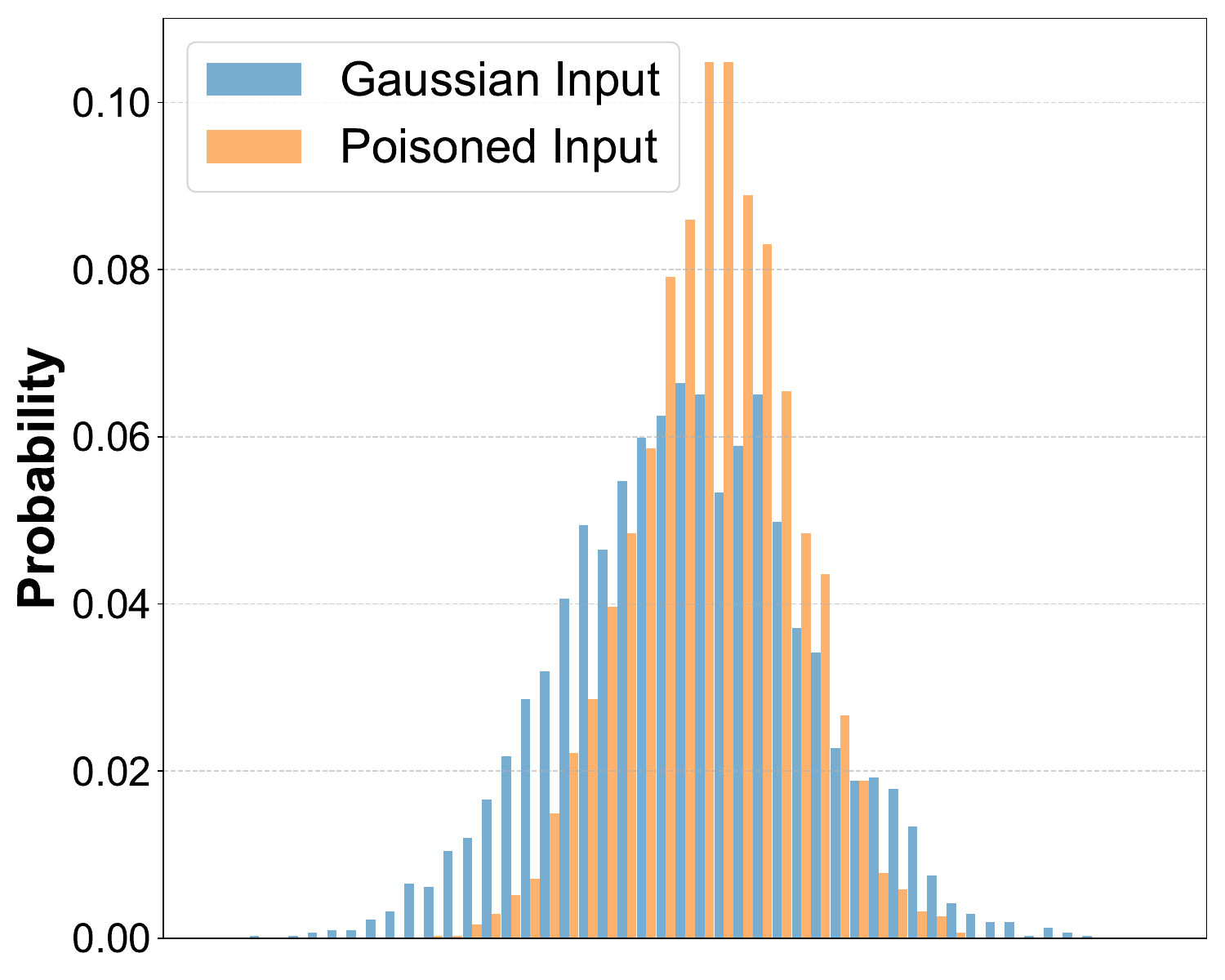}
    \caption{PDD score: 0.174.}
    \label{fig:distribution_blend_b}
  \end{subfigure}
    \begin{subfigure}{0.32\linewidth}
    \includegraphics[width=1\linewidth]{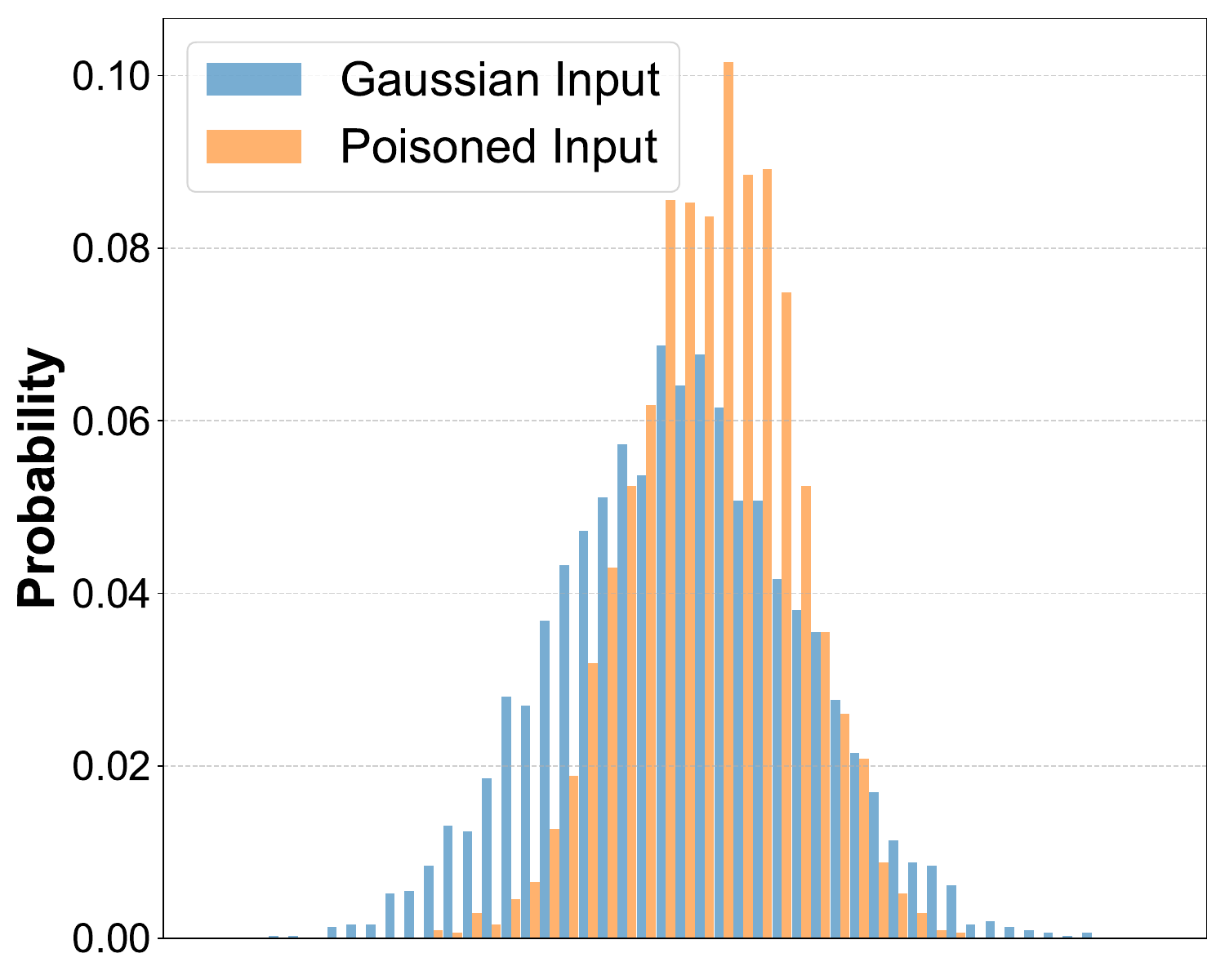}
    \caption{PDD score: 0.167.}
    \label{fig:distribution_blend_c}
    \end{subfigure}

  \caption{Distribution overlap between Gaussian noise and input noise. \textit{\textbf{(Top)}}: Clean noise input (also Gaussian). \textit{\textbf{(Bottom)}}: Poisoned noise input containing Hello Kitty trigger in \cite{chen2023trojdiff}. It is seen that poisoned noise in the prior work exhibits a non-negligible distribution shift, bringing much higher PDD score than benign input.}
  \label{fig:distribution_gaussian_vs_fix}
\end{figure}

\cref{subsec:backdoorDM} shows the feasibility of the backdoor attack on diffusion models via properly diffusing the target distribution and learning to reverse the backdoor generative process. Following this philosophy, some recent works \cite{chen2023trojdiff,chou2023backdoor}  have successfully launched the attack and demonstrated the vulnerability of the backdoored diffusion models with the presence of trigger patterns. However, we argue that the \textit{detectability}, as an important attack performance metric, is not fully considered in the existing studies. More specifically, the embedded trigger patterns used in the state-of-the-art diffusion model backdoor attacks can be effectively detected in a low-cost way.

Our key finding is that the poisoned noise containing the backdoor trigger can be distinguished from the clean Gaussian noise from the lens of data distribution. As illustrated in \cref{fig:distribution_gaussian_vs_fix}, the distributions of two Gaussian noise inputs are highly overlapped; while an obvious distribution shift can be identified when comparing the poisoned noise and the benign Gaussian noise (see \cref{fig:distribution_gaussian_vs_fix}). Such phenomenon implies that the distribution discrepancy between the input and Gaussian noise $\mathcal{N}(0, \mathbf{I})$ can serve as a good marker to detect whether the input is potentially stamped with the backdoor trigger or not. To quantitatively measure this discrepancy, we propose to define a KL divergence-based \cite{kullback1959information} Poisoned Distribution Discrepancy (PDD) score as follows:
\begin{equation}
\begin{aligned}
D(\Tilde{\mathbf{x}}_T) = \texttt{KL}(P_{h(\Tilde{\mathbf{x}}_T)}, P_{h(\mathbf{x}_T)}), \\
\end{aligned}
\label{eqn:kl}
\end{equation}
where $\mathbf{x}_T$ and $\Tilde{\mathbf{x}}_T$ are the clean Gaussian input and the potentially poisoned input, respectively, $h(\cdot)$ denotes the histogram function, $P_{h(\cdot)}$ normalizes the histogram into a probability distribution, and $\texttt{KL}(\cdot, \cdot)$ calculates the KL divergence.

In general, for each potentially poisoned input, we can calculate its PDD score to evaluate its distribution shift from the benign Gaussian noise input. Notice that since even two clean Gaussian noises sampled from the same distribution still have a certain level of distribution discrepancy, such inevitable ``base difference" incurred by the sampling randomness should be considered, and hence it can be empirically calculated as follows:
\begin{equation}
\begin{aligned}
\phi_{Base} = \mathbb{E}_{\mathbf{x}_T} [D({\mathbf{x}}_T)] + 3\sigma_{D({\mathbf{x}}_T)}, 
\end{aligned}
\end{equation}
where $\mathbf{x}_T^{1}, \mathbf{x}_T^{2}, \cdots, \mathbf{x}_T^{N}$ are the collection of clean inputs sampled from Gaussian distribution $\mathcal{N}(0, \textbf{I})$. Also, considering the potential impact of statistical error on false positive rate, the calculation of base discrepancy includes an extra tolerance term (empirically set as $3\sigma_{D({\mathbf{x}}_T)}$), ensuring that most ($>99.8\%$) clean Gaussian noise inputs can be correctly recognized. Then, we can use this base discrepancy as the threshold to detect the backdoor trigger as follows :






\textit{\textbf{PDD-based Trigger Detection.}} \textit{Given an input noise $\Tilde{\mathbf{x}}_T$, it will be detected as poisoned with backdoor trigger if $D(\Tilde{\mathbf{x}}_T) \geq \phi_{Base}$; otherwise it is marked as clean.}

\begin{figure}[t]
    \centering
    \includegraphics[width=1\linewidth]{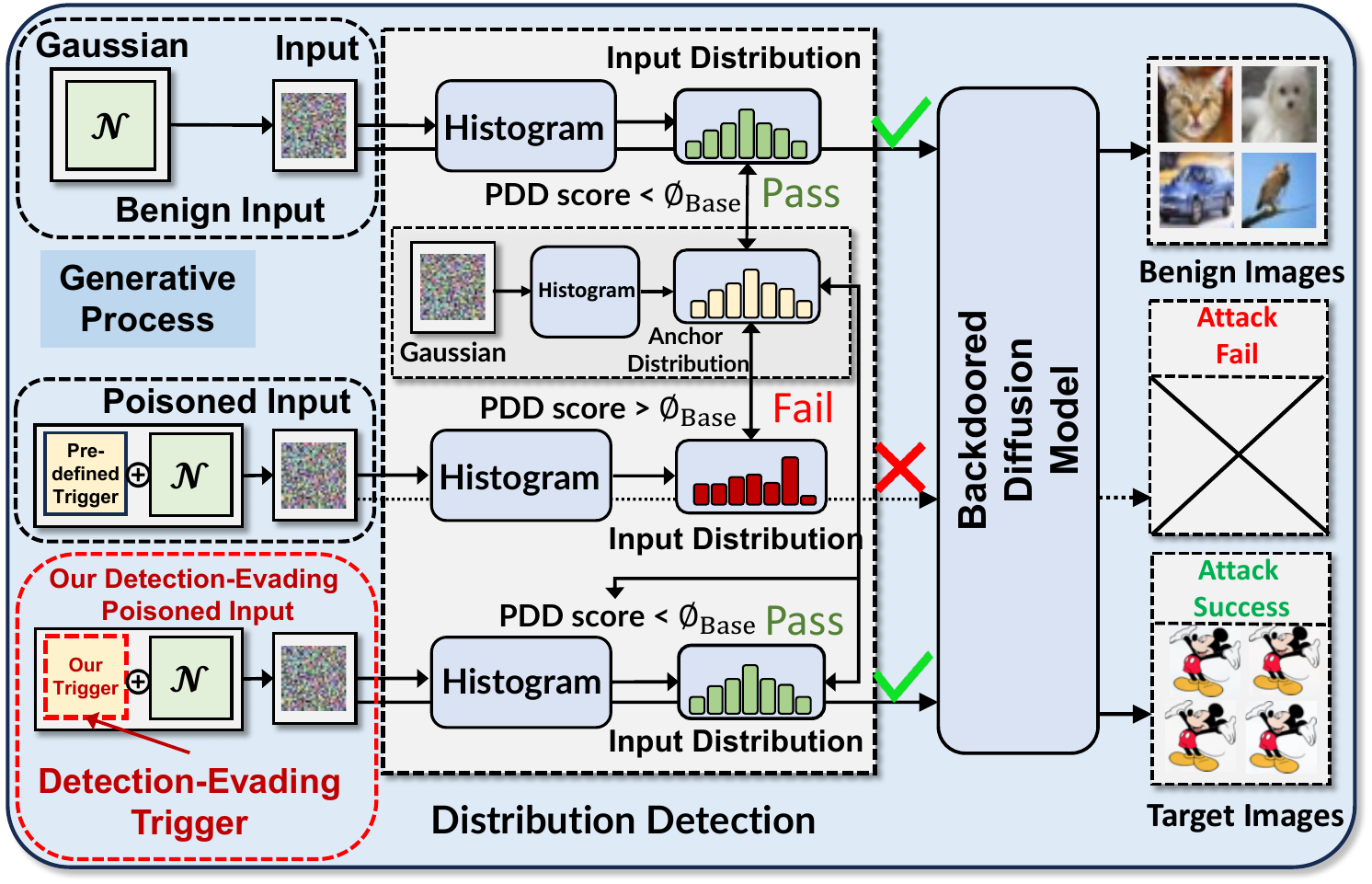}
    \caption{The mechanism of our proposed distribution detection. After calculating the ``anchor" distribution, it can correctly recognize the benign input while effectively identifying the poisoned input designed in the existing backdoored diffusion works, making the attack fail. On the other hand, our proposed detection-evading trigger has a below-threshold PDD score, evading the detection of the distribution detector.}
    \label{fig:defense}
\end{figure}

\cref{fig:defense} illustrates the overall mechanism of the proposed distribution-based trigger detection approach. By preparing a set of clean Gaussian noise to compute the ``anchor" distribution $P_{h(\mathbf{x}_T)}$ and base discrepancy $\phi_{Base}$ as the threshold, the detector can identify the poisoned noise input in a low-cost way. As reported in our empirical evaluations (see \cref{tab:detection_results}), examining distribution shift shows very strong performance for detecting backdoor triggers.

\section{Detection-Evading Backdoor Trigger Design}
\label{sec:attack}




\begin{figure*}[ht]
    \centering
    \includegraphics[width=1\linewidth]{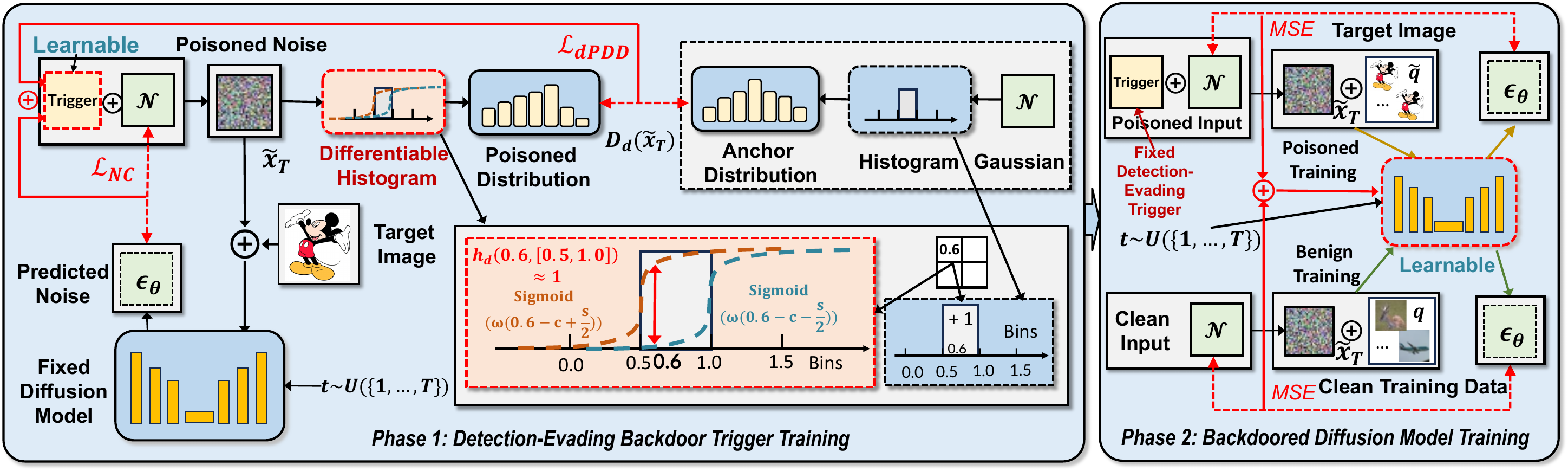}
    \caption{Our proposed two-step training scheme to learn the detection-evading trigger and the corresponding backdoored diffusion model. \textbf{\textit{Phase 1 (Left)}}: Trigger is optimized by PDD loss $\mathcal{L}_{dPDD}$ and NC loss $\mathcal{L}_{NC}$ with the fixed diffusion model. To incorporate an end-to-end training procedure, we utilize the differentiable histogram $h_d(\cdot)$ for calculating $\mathcal{L}_{dPDD}$. \textbf{\textit{Phase 2 (Right)}}: After optimizing the trigger, the diffusion model is updated towards the backdoored training objective with this detection-evading trigger.}
    \label{fig:attack_overall}
\vspace{-2mm}
\end{figure*}

\cref{sec:detection} analyzes the unique characteristics of the backdoor triggers for the diffusion models, and then develops the corresponding detection method. To deepen our understanding, in this section we further study the vulnerability of backdoored diffusion models from the perspective of attackers, exploring stealthy trigger design to evade the distribution-based detection mechanism.





\subsection{Mitigate Distribution Discrepancy}
\label{subsec:pdd}

As analyzed in \cref{sec:detection}, embedding the trigger to the benign Gaussian noise brings the detectable distribution shift. Therefore, in order to make the backdoor trigger undetectable, the PDD score of the poisoned noise $\Tilde{\mathbf{x}}_T$ should be optimized and suppressed below the base discrepancy $\phi_{Base}$ as follows:
\vspace{-2mm}
\begin{equation}
\begin{aligned}
\max_{\boldsymbol{\delta}} \quad
                    & \frac{1}{N} \sum\limits_{i=1}^{i=N} \mathbf{1}(D(\Tilde{\mathbf{x}}_T^i) \leq \phi_{Base}), \\
\end{aligned}
\label{eqn:max_dpr}
\vspace{-2mm}
\end{equation}
where $\mathbf{1}(\cdot)$ is the indicator function, and $\Tilde{\mathbf{x}}_T^i$ is one sample of poisoned noise input. Notice that here in order to mitigate the sampling error, the optimization of the backdoor trigger $\boldsymbol{\delta}$ is based on the evaluation of embedding $\boldsymbol{\delta}$ to $N$ benign Gaussian noise inputs $\mathbf{x}_T^i$. Then, the trigger can be learned via minimizing the following \textit{PDD loss}:
\begin{equation}
\begin{aligned}
\mathcal{L}_{PDD}({\boldsymbol{\delta}}) = \mathbb{E}_{\Tilde{\mathbf{x}}_T}[\max(D(\Tilde{\mathbf{x}}_T) - \phi_{Th}, 0)], 
\end{aligned}
\label{eqn:loss_ddp}
\end{equation}
where $\phi_{Th}$ is a pre-set threshold even smaller than $\phi_{Base}$, ensuring that after the training, the optimized PDD score can be optimized as being below $\phi_{Base}$ in a very probable way.

\noindent \textbf{Make Trigger Learning Differentiable.} In general, optimizing the PDD loss can be realized by using a gradient-based method such as stochastic gradient descent (SGD). However, as described in \cref{eqn:kl}, the calculation of PDD score $D(\Tilde{\mathbf{x}}_T)$ is involved with the non-differentiable histogram function $h(\cdot)$, preventing the differentiable learning of stealthy trigger. To address this problem, we propose to approximate the original histogram function to a differentiable format $h_d(\cdot)$. Here, the key idea is to use the dual logistic functions as a closed surrogate for the histogram function (see \cref{fig:attack_overall}). More specifically, the differentiable histogram is calculated as:
\begin{equation}
\begin{aligned}
h_d(\mathbf{x}, B_{i}) = &  \sum\limits_{\mathbf{x}} ((1 + e^{-\omega(\mathbf{x}-c_{i}+\frac{s_i}{2})})^{-1} \\
                        & - (1 + e^{-\omega(\mathbf{x}-c_{i}-\frac{s_i}{2})})^{-1}),
\end{aligned}
\label{eqn:diff_histogram}
\end{equation}
where $\omega$ controls the smoothness of the histogram, $B_i$ denotes the $i$-th bin in the histogram, and $c_{i}$ and $s_i$ represent the center and width of each bin, respectively. Then, the differentiable version of PDD score $D_d(\cdot)$ and loss used can be calculated as:
\begin{equation}
\begin{aligned}
D_d(\Tilde{\mathbf{x}}_T) &= \texttt{KL}(P_{h_d(\Tilde{\mathbf{x}}_T)}, P_{h_d(\mathbf{x}_T)}), \\
\mathcal{L}_{dPDD}({\boldsymbol{\delta}}) &= \mathbb{E}_{\Tilde{\mathbf{x}}_T}[\max(D_d(\Tilde{\mathbf{x}}_T) - \phi_{Th}, 0)]. 
\end{aligned}
\label{eqn:loss_pdd_2}
\end{equation}

\noindent \textbf{Two-Step Learning Procedure.} With the availability of differentiable PDD loss, the backdoored diffusion model and the corresponding detection-evading can be learned in an end-to-end manner. As shown in \cref{fig:attack_overall}, we first fix the to-be-backdoored diffusion model and optimize the trigger by using PDD loss and NC loss (described in \cref{subsec:NC}). After obtaining the stealthy trigger exhibiting low distribution discrepancy, we then fix this trigger and use it to generate poisoned input noise, facilitating the poison training for the backdoored diffusion model.

\subsection{Noise Consistency Optimization}
\label{subsec:NC}

As shown in \cref{fig:attack_overall}, in the trigger training phase, the \textit{noise consistency loss} ($\mathcal{L}_{NC}$), which measures the discrepancy between the benign Gaussian noise input and the predicted noise $\boldsymbol{\epsilon}_{\theta}$, is also used to guide the optimization of backdoor trigger $\boldsymbol{\delta}$. More specifically, the NC loss is defined and calculated as follows:
\begin{equation}
\begin{aligned}
\mathcal{L}_{NC}(\boldsymbol{\delta}, \Tilde{\mathbf{x}}_0) = & \mathbb{E}_{t, \Tilde{\mathbf{x}}_0, \boldsymbol{\epsilon}} [\| \boldsymbol{\epsilon} - \boldsymbol{\epsilon}_{\theta}(\sqrt{\overline{\alpha}_{t}}\Tilde{\mathbf{x}}_0 \\
& + \sqrt{1-\overline{\alpha}_t}\boldsymbol{\mu}_{\boldsymbol{\delta}} + \sqrt{1-\overline{\alpha}_{t}}\gamma\boldsymbol{\epsilon}, t) \|^{2}], \\
\end{aligned}
\label{eqn:loss_nc}
\end{equation}
where $\Tilde{\mathbf{x}}_0, t, \boldsymbol{\epsilon}, \boldsymbol{\epsilon}_{\theta}$ are with the same setting in \cref{eqn:backdoor_dm_objective}. Here, the use of NC loss is motivated by the following design philosophy: Because the backdoored model training process (Phase 2) will use the exactly same discrepancy to update the learnable model (see \cref{fig:attack_overall}), pre-optimize this loss in the trigger learning phase can provide better initialization and hence potentially improve both benign and attack performance. To be specific, with lower discrepancy between $\boldsymbol{\epsilon}$ and $\boldsymbol{\epsilon}_{\theta}$ for the poisoned training part, slight update from the original benign model may be already sufficient for fitting poisoned data samples, and hence the updated model, which is backdoored but closed to the original benign one, can probably perform well with the presence of benign inputs. Meanwhile, with the lower NC loss as the initialization, it is more likely to bring the poisoned training \cref{fig:attack_overall} to a better-optimized point after the same number of epochs, thereby improving the attack performance with the poisoned noise inputs. Notice that such a hypothesis has been verified in our empirical evaluations reported in \cref{sec:ablationstudies}. Algorithm \ref{alg:overall} describes the overall 2-step training procedure, including using NC loss.

\begin{algorithm}[ht]
    \footnotesize
    \caption{The Proposed 2-Step Training Scheme}
    \label{alg:overall}
        \textbf{Input:} Clean dataset $q(\mathbf{x}_0)$, backdoor target dataset $\Tilde{q}(\Tilde{\mathbf{x}}_0)$, pre-trained benign diffusion model $\theta$, scaling factor $\tau$, propositional factor $\gamma$, threshold $\phi_{Th}$, trigger learning rate $\eta_{t}$, model learning rate $\eta_{d}$. 
        \par
        \textbf{Output:} Detection-evading  trigger $\boldsymbol{\delta}$, backdoored diffusion model $\theta_{bd}$. \par
    \begin{algorithmic}[1]
    \STATE \textit{\textcolor{teal}{ Phase 1: Detection-Evading Backdoor Trigger Training}}
    \STATE $\boldsymbol{\delta} \leftarrow \texttt{random} (\boldsymbol{\delta} \texttt{.shape})$
    \REPEAT
        \STATE $\Tilde{\mathbf{x}}_0\sim \Tilde{q}(\Tilde{\mathbf{x}}_0), {t} \sim \text{Uniform}{(\{ 1,\cdots,T\})}, {\boldsymbol{\epsilon}}\sim\mathcal{N}(0, \mathbf{I})$
        \STATE $\Tilde{\mathbf{x}}_T \sim \mathcal{N}(\boldsymbol{\mu}_{\boldsymbol{\delta}}, \gamma \boldsymbol{\epsilon})$ 
        \textcolor{brown}{\COMMENT{\textit{Sample poisoned input noises}}}
        \STATE $\mathcal{L}_{dPDD}({\boldsymbol{\delta}}) = \mathbb{E}_{\Tilde{\mathbf{x}}_T}[\max(D_d(\Tilde{\mathbf{x}}_T) - \phi_{Th}, 0)]$ via Eq. \ref{eqn:loss_pdd_2}
        \STATE $\mathcal{L}_{NC}(\boldsymbol{\delta}; \Tilde{\mathbf{x}}_0) = \mathbb{E}_{\Tilde{\mathbf{x}}_0, {t}, {\boldsymbol{\epsilon}}}[\| \boldsymbol{\epsilon} - \boldsymbol{\epsilon}_\theta(\Tilde{\mathbf{x}}_0, {t}, {\boldsymbol{\epsilon}}, \boldsymbol{\delta}) \|]$ via Eq. \ref{eqn:loss_nc}
        \STATE $\mathcal{L}(\boldsymbol{\delta}; \Tilde{\mathbf{x}}_0) = \mathcal{L}_{NC}(\boldsymbol{\delta}; \Tilde{\mathbf{x}}_0) + \tau\mathcal{L}_{dPDD}(\boldsymbol{\delta})$ 
        \textcolor{brown}{\COMMENT{\textit{Overall trigger loss}}}
        \STATE $\boldsymbol{\delta} \leftarrow \boldsymbol{\delta} - \eta_{t} \nabla_{\boldsymbol{\delta}}$ $\mathcal{L}(\boldsymbol{\delta}; \Tilde{\mathbf{x}}_0)$ 
        \textcolor{brown}{\COMMENT{\textit{Updating trigger $\delta$}}}

    \UNTIL converged
    \STATE \textit{\textcolor{teal}{ Phase 2: Backdoored Diffusion Model Training}}
    \STATE $\theta_{bd} \leftarrow \theta$ \textcolor{brown}{\COMMENT{\textit{Loading pre-trained benign diffusion model}}}
    \REPEAT 
        \STATE ${\mathbf{x}_0} \sim q({\mathbf{x}_0}), t \sim \text{Uniform}{(\{ 1,\cdots,T\})}, \boldsymbol{\epsilon}\sim\mathcal{N}(0, \mathbf{I})$
        \STATE $\Tilde{\mathbf{x}}_0\sim \Tilde{q}(\Tilde{\mathbf{x}}_0), \Tilde{t} \sim \text{Uniform}{(\{ 1,\cdots,T\})}, \Tilde{\boldsymbol{\epsilon}}\sim\mathcal{N}(0, \mathbf{I})$
        \STATE $\mathcal{L}_{c}(\theta_{bd}) = \mathbb{E}_{{\mathbf{x}}_0, {t}, {\boldsymbol{\epsilon}}}[\| \boldsymbol{\epsilon} - \boldsymbol{\epsilon}_{\theta_{bd}}(\mathbf{x}_0, t, \boldsymbol{\epsilon}) \|]$ via Eq. \ref{eqn:dm_objective}
        \textcolor{brown}{\COMMENT{\textit{Benign}}}
        \STATE $\mathcal{L}_{p}(\theta_{bd}) = \mathbb{E}_{\Tilde{\mathbf{x}}_0, \Tilde{t}, \Tilde{\boldsymbol{\epsilon}}}[\| \boldsymbol{\epsilon} - \boldsymbol{\epsilon}_{\theta_{bd}}(\Tilde{\mathbf{x}}_0, \Tilde{t}, \Tilde{\boldsymbol{\epsilon}}, \boldsymbol{\delta}) \|]$ via Eq. \ref{eqn:backdoor_dm_objective}
        \textcolor{brown}{\COMMENT{\textit{Poison}}}
        \STATE $\mathcal{L}(\theta_{bd}) = \mathcal{L}_{c}(\theta_{bd})+\mathcal{L}_{p}(\theta_{bd})$
        \textcolor{brown}{\COMMENT{\textit{Backdoored model loss}}}
        \STATE $\theta_{bd} \leftarrow \theta_{bd} - \eta_{d} \nabla_{\theta_{bd}} \mathcal{L}(\theta_{bd})$ 
        \textcolor{brown}{\COMMENT{\textit{Updating diffusion model}}}
    \UNTIL converged
    \end{algorithmic} 
  \end{algorithm}

\section{Experiments}
\label{sec:experiments}


\subsection{Settings}

\begin{table}[t]
  \setlength{\tabcolsep}{1pt}
  \centering
  \scalebox{0.8}{
  \begin{tabular}{lcc|cc|cc}
    \toprule
    
    \multirow{4}{*}{\makecell{\textbf{Attack} \\ \textbf{Mode}}} & \multirow{4}{*}{\makecell{\textbf{Average} \\ \textbf{PDD} \\ \textbf{Score}}} & \multirow{4}{*}{\makecell{\textbf{Detection} \\ \textbf{Rate} \\ \textbf{(\%)}}} & \multicolumn{2}{c|}{\textbf{DDPM}} & \multicolumn{2}{c}{\textbf{DDIM}} \\
    
    \cline{4-7}
    & & & \multirow{3}{*}{\makecell{\textbf{ASR (\%)} \\ \textbf{w/o} \\ \textbf{Detection}}} & \multirow{3}{*}{\makecell{\textbf{ASR} (\%) \\ \textbf{w/} \\ \textbf{Detection}}} & \multirow{3}{*}{\makecell{\textbf{ASR (\%)} \\ \textbf{w/o} \\ \textbf{Detection}}} & \multirow{3}{*}{\makecell{\textbf{ASR (\%)} \\ \textbf{w/} \\ \textbf{Detection}}} \\ 
    
    & & & & & & \\
    & & & & & & \\
    \midrule
        \multicolumn{7}{c}{CIFAR-10} \\
    \hhline{-|-|-|-|-|-|-|}
    Category & 0.183 & \textbf{100.0} & 90.1 & \textbf{0.0} & 87.30 & \textbf{0.0} \\
    Instance & 0.183 & \textbf{100.0} & 100.0 & \textbf{0.0} & 100.0 & \textbf{0.0}     \\
    \midrule
        \multicolumn{7}{c}{CelebA} \\
    \hhline{-|-|-|-|-|-|-|}
    Category & 0.165 & \textbf{100.0} &  96.9 & \textbf{0.0} & 95.4 & \textbf{0.0} \\
    Instance & 0.165 & \textbf{100.0} & 100.0 & \textbf{0.0} & 100.0 & \textbf{0.0} \\

    \bottomrule
  \end{tabular}
  }
  \caption{The effectiveness of the proposed distribution-based detection method for detecting the backdoor trigger used in \cite{chen2023trojdiff}. The PDD score $D_d(\Tilde{\mathbf{x}}_T) \gg \phi_{Base}=0.067$ (for CIFAR-10 dataset) and $D_d(\Tilde{\mathbf{x}}_T) \gg \phi_{Base}=0.016$ (for CelebA dataset), \textbf{making the detection rate reach to $\textbf{100\%}$ and ASR drop to $\textbf{0\%}$}.}
  \label{tab:detection_results}
  \vspace{-4mm}
\end{table}

\noindent \textbf{Datasets, Models and Attack Setting.} We evaluate the performance of the proposed detection method and the detection-evading trigger for DDPM \cite{ho2020denoising} and DDIM \cite{song2020denoising} diffusion models on CIFAR-10 (32 $\times$ 32) \cite{krizhevsky2009learning} and CelebA (64 $\times$ 64) \cite{liu2015deep} datasets. The pre-trained models of CIFAR-10 and CelebA datasets are from \texttt{pesserpytorch$/$diffusion} and \texttt{ermongroup/ddim}. Two types of backdoor attack models are considered in the experiments: generate an image belonging to a specific class (referred to as ``category mode") and generate a specific image (referred to as ``instance mode"). Following the settings in \cite{chen2023trojdiff},  we choose the horse in the CIFAR-10 dataset and faces with heavy makeup, mouth slightly open and smiling in the CelebA dataset as the target class in the category mode. The Micky Mouse is selected as the target image when the backdoor attack is launched in the instance mode. The ``Hello Kitty" and ``Glass" images are set as fixed triggers in experiments of TrojDiff \cite{chen2023trojdiff} and Baddiffusion \cite{chou2023backdoor}, respectively. 

\noindent \textbf{Training Configurations.} When training the detection-evading trigger (Phase 1), an \texttt{Embedding} layer with the same shape of input noise ($3\times32\times32$ for CIFAR-10 dataset, and $3\times64\times64$ for CelebA dataset) is used for trigger learning with $\gamma=0.6$. The threshold is set as $\phi_{Th}=0.01$ and $\phi_{Th}=0.005$ for the CIFAR-10 and CelebA datasets, respectively. The training process adopts Adam optimizer \cite{KingBa15} with 50k training steps, $2\times10^{-3}$ learning rate and scaling factor $\tau$ as $10^4$. 
After that, during the training procedure for the backdoored diffusion model (Phase 2), we follow the standard training procedure using Adam optimizer, $2 \times 10^{-4}$ learning rate, batch size as 256, and 100k training steps. Also, the number of bins is set as 50 for both regular histogram $h(\cdot)$ and differentiable histogram $h_d({\cdot})$. The smoothness parameter is set as $\omega=6$ for the Sigmoid function in $h_d({\cdot})$ to approximate the step function and histogram $h(\cdot)$. All the experiments are conducted on NVIDIA RTX A6000 GPUs.

\noindent \textbf{Evaluation Metrics.} The benign performance is evaluated on 50K samples via measuring Frechet Inception Distance (FID) \cite{heusel2017gans}, which reveals the similarity between two sets of images. A lower FID score indicates the higher quality of the generated images. The attack performance is evaluated on 10K samples in terms of Attack Success Rate (ASR). When the attack mode is set as ``category" and ``instance", ASR is measured as the ratio of the generated images being classified into the target class and being the same as the target image, respectively. Specifically, when the attack is launched in the ``instance" mode, we also measure the Mean Square Error (MSE) to examine the difference between the target image and the generated images. For the image sampling, we follow the standard strategy by setting $\eta = 1$ and $S = 1000$ and $\eta = 0$ and $S = 100$ in DDPM \cite{song2020denoising} and DDIM \cite{ho2020denoising}, respectively.


\begin{table*}[t]
  \centering
  \scalebox{0.8}{
  \begin{tabular}{c|cc|cc|cc|c|cc|c}
    \toprule
    
    \multirow{4}{*}{\makecell{\textbf{Attack} \\ \textbf{Mode}}} & \multirow{4}{*}{\textbf{Method}} & \multirow{4}{*}{\makecell{\textbf{Trigger} \\ \textbf{Type}}} & \multirow{4}{*}{\makecell{\textbf{PDD} \\ \textbf{Score}}} & \multirow{4}{*}{\makecell{\textbf{Detection} \\ \textbf{Pass Rate} \\ \textbf{(\%)}}} & \multicolumn{3}{c|}{\textbf{DDPM}} & \multicolumn{3}{c}{\textbf{DDIM}}\\
    \cline{6-11}
    
    & & & & & \multicolumn{2}{c|}{\textbf{Benign}} & \textbf{Attack} & \multicolumn{2}{c|}{\textbf{Benign}} & \textbf{Attack} \\
    \cline{6-11}

    & & & & & \multirow{2}{*}{\makecell{\textbf{FID}  $\downarrow$ }} & \multirow{2}{*}{\makecell{$\Delta$ \\ \textbf{FID}}} & \multirow{2}{*}{\makecell{\textbf{ASR (\%) w/} \\ \textbf{Detection} $\uparrow$ }} & \multirow{2}{*}{\makecell{\textbf{FID}  $\downarrow$ }} & \multirow{2}{*}{\makecell{$\Delta$ \\ \textbf{FID}}} & \multirow{2}{*}{\makecell{\textbf{ASR (\%) w/} \\ \textbf{Detection $\uparrow$} }} \\

    & & & & & & & & & \\
    
    \midrule
    \multicolumn{11}{c}{CIFAR-10} \\  
    \midrule
    
    None & Benign Baseline & None & 0.031$\pm$0.012 & 99.8 & 4.60 & 0 & {0} & 4.25 & 0 & {0}\\
    
    \midrule
    & Trojdiff \cite{chen2023trojdiff} & Fixed & 0.183$\pm$0.012 & 0.0 & \multirow{1}{*}{4.74} & \multirow{1}{*}{0.14} & 0.0 & \multirow{1}{*}{4.47} & \multirow{1}{*}{0.22} & 0.0 \\
    \hhline{~|-|-|-|-|-|-|-|-|-|-|}

     \rowcolor{gray!10} \cellcolor{white} \multirow{-2}{*}{Category} & \textbf{DisDet(Ours)} & \textbf{Learnable} & \textbf{0.025$\pm$0.007} & \textbf{99.9} & {\textbf{4.44}} & {\textbf{-0.16}} & \textbf{82.0} & {\textbf{4.29}} & {\textbf{0.04}} & \textbf{80.1} \\
    
    \midrule
      & Trojdiff \cite{chen2023trojdiff} & Fixed & 0.183$\pm$0.012 & 0.0 & {4.59} & {-0.01} & 0.0 & {4.47} &{0.22} & 0.0 \\
    \hhline{~|-|-|-|-|-|-|-|-|-|-|}
        &  Baddiffusion$^\ast$ \cite{chou2023backdoor} & Fixed & 0.269$\pm$0.015 & 0.0 & {4.52} & {-0.08} & 0.0 & {4.43} &{0.18} & 0.0 \\
    \hhline{~|-|-|-|-|-|-|-|-|-|-|}
    
     \rowcolor{gray!10} \cellcolor{white} & & & & & & & \textbf{99.9} & & & \textbf{99.9} \\
     \rowcolor{gray!10} \cellcolor{white} \multirow{-4}{*}{Instance} & \multirow{-2}{*}{\textbf{DisDet(Ours)}} & \multirow{-2}{*}{\textbf{Learnable}} & \multirow{-2}{*}{\textbf{0.025$\pm$0.007}} & \multirow{-2}{*}{\textbf{99.9}} & \multirow{-2}{*}{{\textbf{4.39}}} & \multirow{-2}{*}{{\textbf{-0.21}}} & (\textbf{MSE: 7.64e-6}) & \multirow{-2}{*}{{\textbf{4.38}}} & \multirow{-2}{*}{{\textbf{0.13}}} & \textbf{(MSE: \textbf{4.19e-5})} \\

    
    \midrule
    \multicolumn{11}{c}{CelebA} \\
    \midrule

    \multirow{1}{*}{None} & Benign Baseline & None & 0.007$\pm$0.003 & 99.8 & 5.88 & 0 & {0} & 6.29 & 0 & {0} \\
    
    \midrule
      & Trojdiff \cite{chen2023trojdiff} & Fixed  & 0.165$\pm$0.006 & 0.0 & {5.44} & {-0.44} & 0.0 & {5.40} & {-0.89} & 0.0 \\
   \hhline{~|-|-|-|-|-|-|-|-|-|-|} 
    \rowcolor{gray!10} \cellcolor{white} \multirow{-2}{*}{Category} & \textbf{DisDet(Ours)} & \textbf{Learnable}  & \textbf{0.007$\pm$0.003} & \textbf{99.8} & {\textbf{5.83}} & {\textbf{-0.05}} & \textbf{85.9} & {\textbf{5.94}} & {\textbf{-0.35}} & \textbf{85.2} \\
    \midrule
    & Trojdiff \cite{chen2023trojdiff} & Fixed  & 0.165$\pm$0.006 & 0.0 & {5.62} & {-0.26} & 0.0 & {5.93} & {-0.36} & 0.0 \\
       \hhline{~|-|-|-|-|-|-|-|-|-|-|}
    & Baddiffusion$^\ast$ \cite{chou2023backdoor} & Fixed  & 0.260$\pm$0.007 & 0.0 & {5.73} & {-0.15} & 0.0 & {5.98} & {-0.31} & 0.0 \\
   \hhline{~|-|-|-|-|-|-|-|-|-|-|}
    \rowcolor{gray!10} \cellcolor{white} & & & & & & & \textbf{99.8} & & & \textbf{99.8} \\

    \rowcolor{gray!10} \cellcolor{white} \multirow{-4}{*}{Instance} & \multirow{-2}{*}{\textbf{DisDet(Ours)}} & \multirow{-2}{*}{\textbf{Learnable}}  & \multirow{-2}{*}{\textbf{0.007$\pm$0.003}} & \multirow{-2}{*}{\textbf{99.8}} & \multirow{-2}{*}{\textbf{5.80}} & \multirow{-2}{*}{\textbf{-0.08}} & \textbf{(MSE: 1.52e-3)} & \multirow{-2}{*}{\textbf{5.85}} & \multirow{-2}{*}{{\textbf{-0.44}}} & \textbf{(MSE: 1.70e-3)} \\

    \bottomrule
  \end{tabular}
  }
  \caption{Performance of the proposed backdoored diffusion model using the detection-evading learnable trigger. The lower FID and higher ASR indicate better benign and attack performance, respectively. For ``Instance" mode, MSE between the generated and target images is also reported to show the high attack performance. ``$\ast$" denotes we reproduce the experiments using ``Glasses" \cite{chou2023backdoor} as the trigger and ``Micky Mouse" as the target image.}
  \label{tab:main_results}
  \vspace{-4mm}
\end{table*}

\subsection{Evaluation Results}

\noindent \textbf{CIFAR-10 Dataset.} \Cref{tab:main_results} shows the benign and attack performance on the CIFAR-10 dataset. 
When the backdoor attack is launched in the ``category" mode, the average PDD score of the poisoned noise generated from the trigger designed in \cite{chen2023trojdiff} is 0.183. This score is significantly higher than the base discrepancy $\phi_{Base}=0.067$, making the attack can be easily detected with ASR as 0. In contrast, our detection-evading trigger is learned to exhibit a very low PDD score of 0.025, making the attack very undetectable (nearly $100\%$ detection pass rate) with high ASR (more than $80\%$). Meanwhile, it enjoys good benign performance with even lower FID than the baseline (originally non-backdoored case). In other words, with the presence of benign input, the images generated by our backdoored model have even higher quality than the ones generated by the diffusion model without backdoor injection. Similarly, in the backdoor attack mode ``instance", our approach also shows much better benign and attack performance than the prior works. 
In particular, the MSE between the generated images and the original target image (Mickey Mouse) is very small (7.64e-6 and 4.19e-5 for DDPM and DDIM, respectively), indicating the effectiveness of the attack. 



\noindent \textbf{CelebA Dataset.} As shown in \Cref{tab:main_results}, our optimized trigger is effectively stealthy to the distribution detector and achieves higher attack performance than the prior works. More specifically, with $\phi_{Base}$ as 0.016, the average PDD score of the poisoned noise in \cite{chen2023trojdiff} ( ``category" attack mode) is 0.165, bringing an ASR of 0 since all the triggers will be detected. On the other hand, the average PDD score of our detection-evading trigger is only 0.007, and hence it is very undetectable to the distribution detector, bringing high ASR (more than $85\%$). Meanwhile, the benign performance of our solution is good with even lower FID than the baseline design. Similarly, in the `instance" attack mode, our method enjoys a low average PDD score of 0.007 and also FID reduction as compared to the baseline case, demonstrating high attack and benign performance. 

\begin{figure}[t]
    \centering
    \includegraphics[width=1\linewidth]{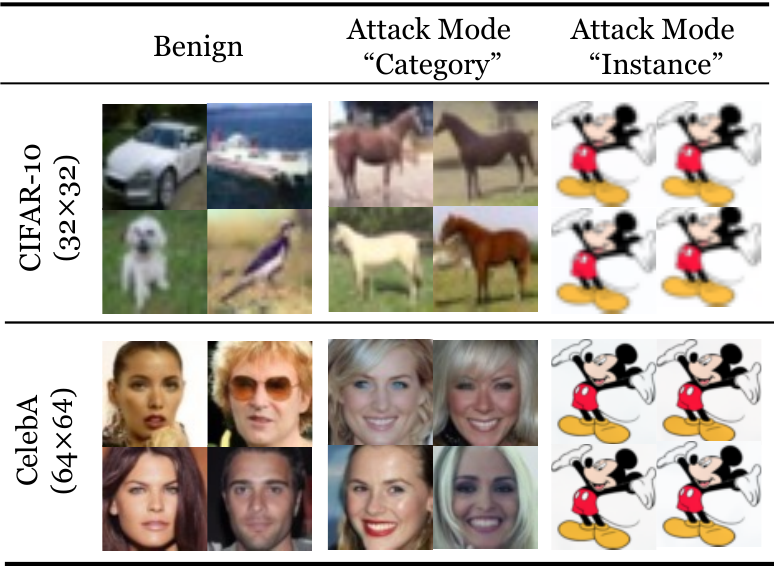}
    \caption{Generated images from our backdoored diffusion model. For CIFAR-10, the target class is ``horse"; for CelebA, the target class includes faces characterized by ``heavy makeup, smiling, and a slightly open mouth". The target image is a ``Michy Mouse".}
    \label{fig:attack_vis}
\vspace{-10mm}
\end{figure}

\noindent \textbf{Visualization.} \cref{fig:attack_vis} illustrates some of the generated images from our backdoored diffusion model with benign and poisoned noise inputs. It is seen that our approach is very effective in both benign and attack scenarios. 

\section{Ablation Studies}
\label{sec:ablationstudies}

\begin{figure}[t]
  \centering
  \begin{subfigure}{0.49\linewidth}
    \includegraphics[width=1\linewidth]{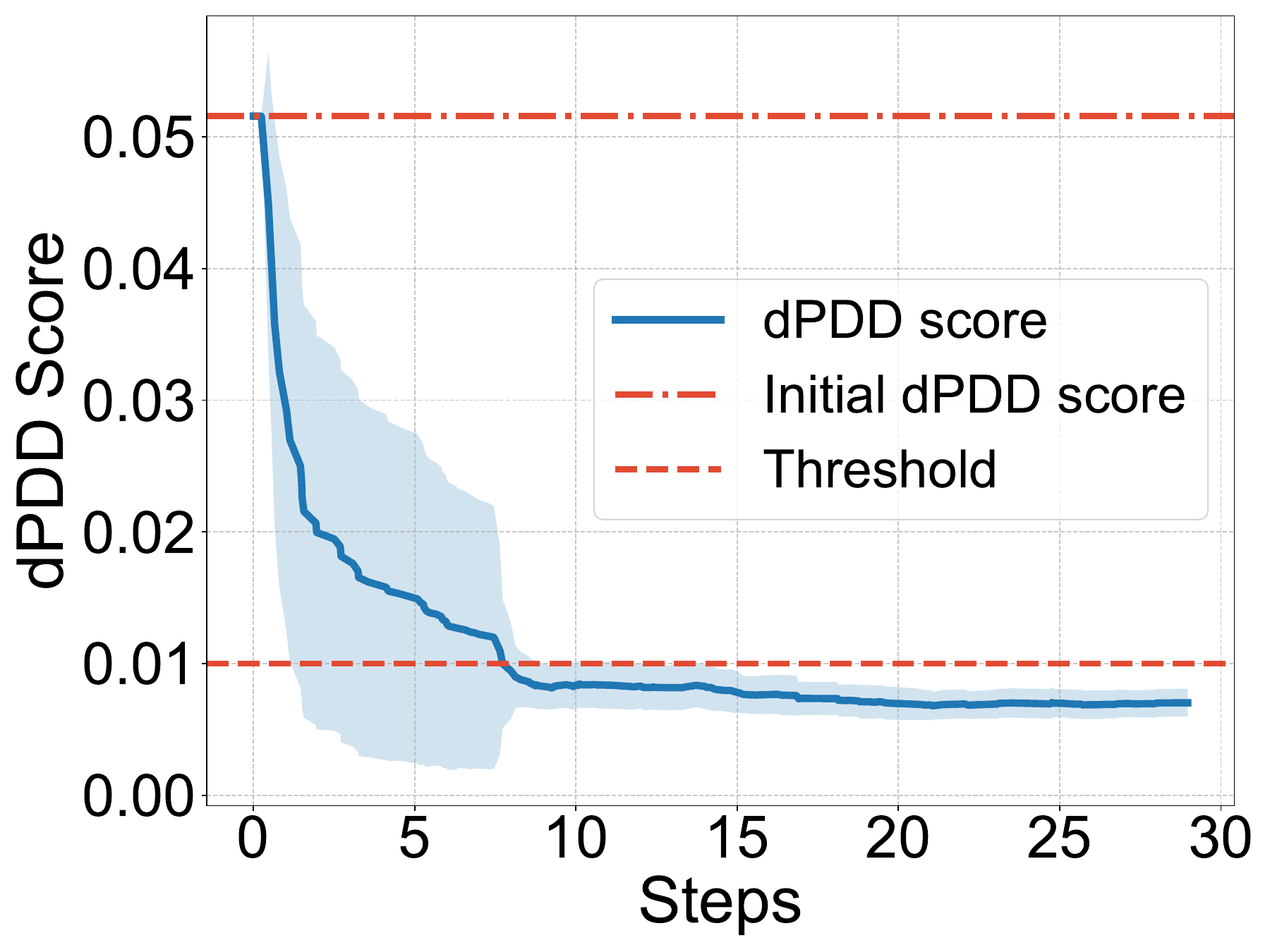}
    \caption{Attack mode: ``Category".}
    \label{fig:effect_kl_a}
  \end{subfigure}
  \hfill
  \begin{subfigure}{0.49\linewidth}
    \includegraphics[width=1\linewidth]{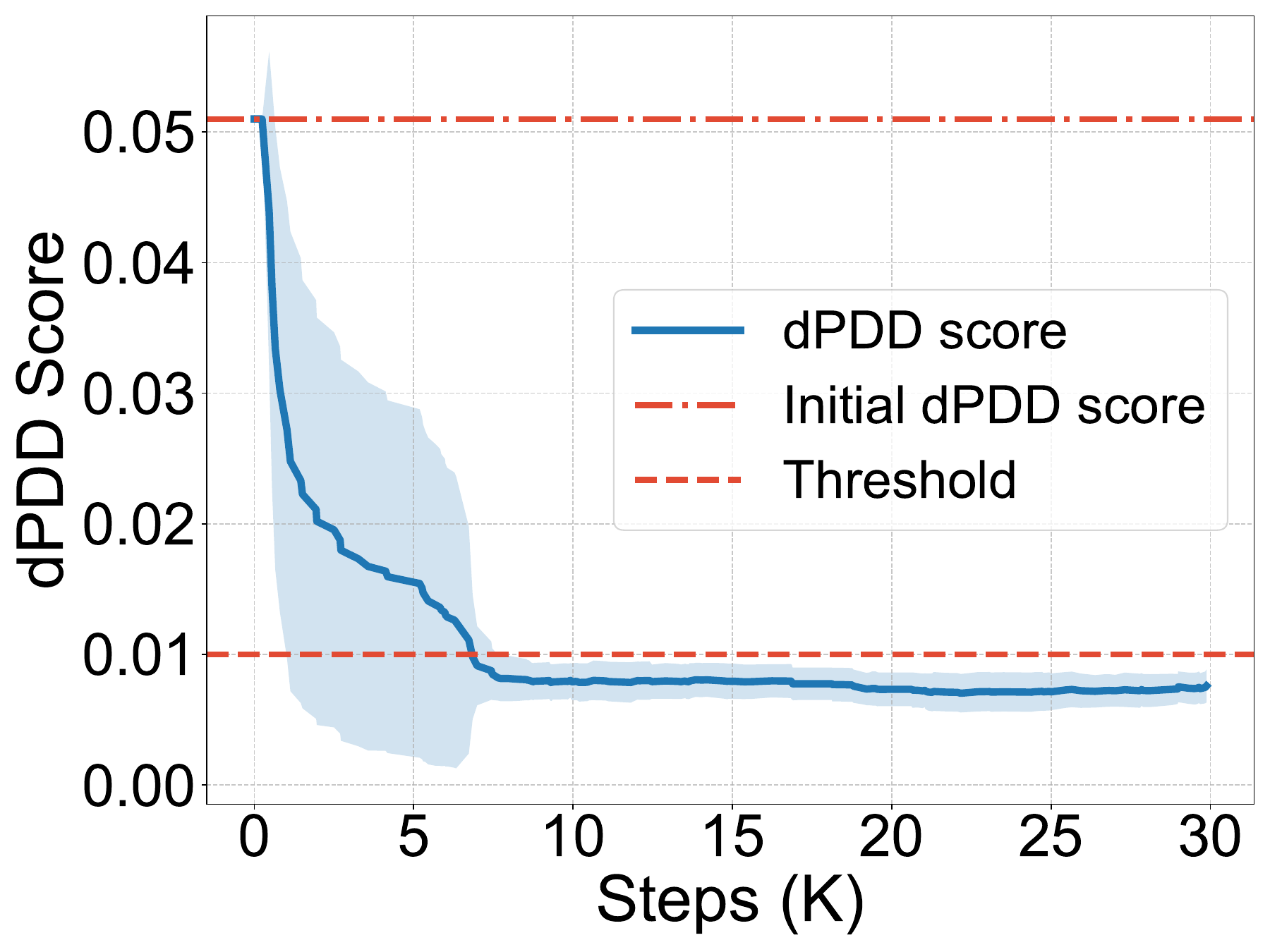}
    \caption{Attack mode: ``Instance".}
    \label{fig:effect_kl_b}
  \end{subfigure}
  \caption{Curve of differentiable PDD score $D_d(\Tilde{\mathbf{x}}_T)$ when the trigger is trained with the PDD loss $\mathcal{L}_{dPDD}$ on the CIFAR-10 dataset. $D_d(\Tilde{\mathbf{x}}_T)$ steadily decreases and reaches below threshold.}
  \label{fig:effect_kl}
  \vspace{-4mm}
\end{figure}

\begin{figure}[t]
  \centering
    \begin{subfigure}{0.49\linewidth}
    \includegraphics[width=1\linewidth]{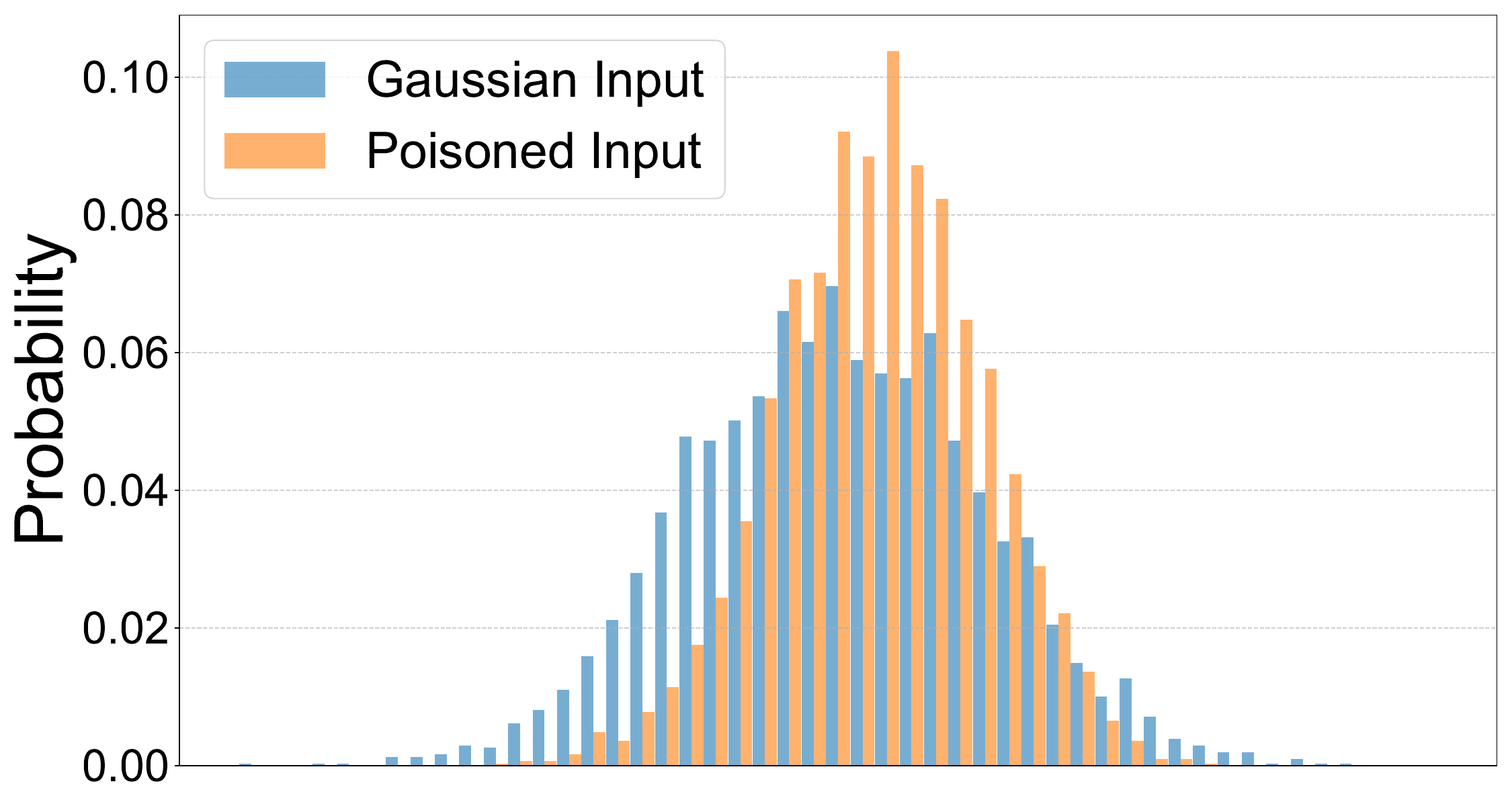}
    \caption{PDD score: 0.179.}
    \label{fig:distribution_ours_a}
  \end{subfigure}
  \hfill
  \begin{subfigure}{0.49\linewidth}
    \includegraphics[width=1\linewidth]{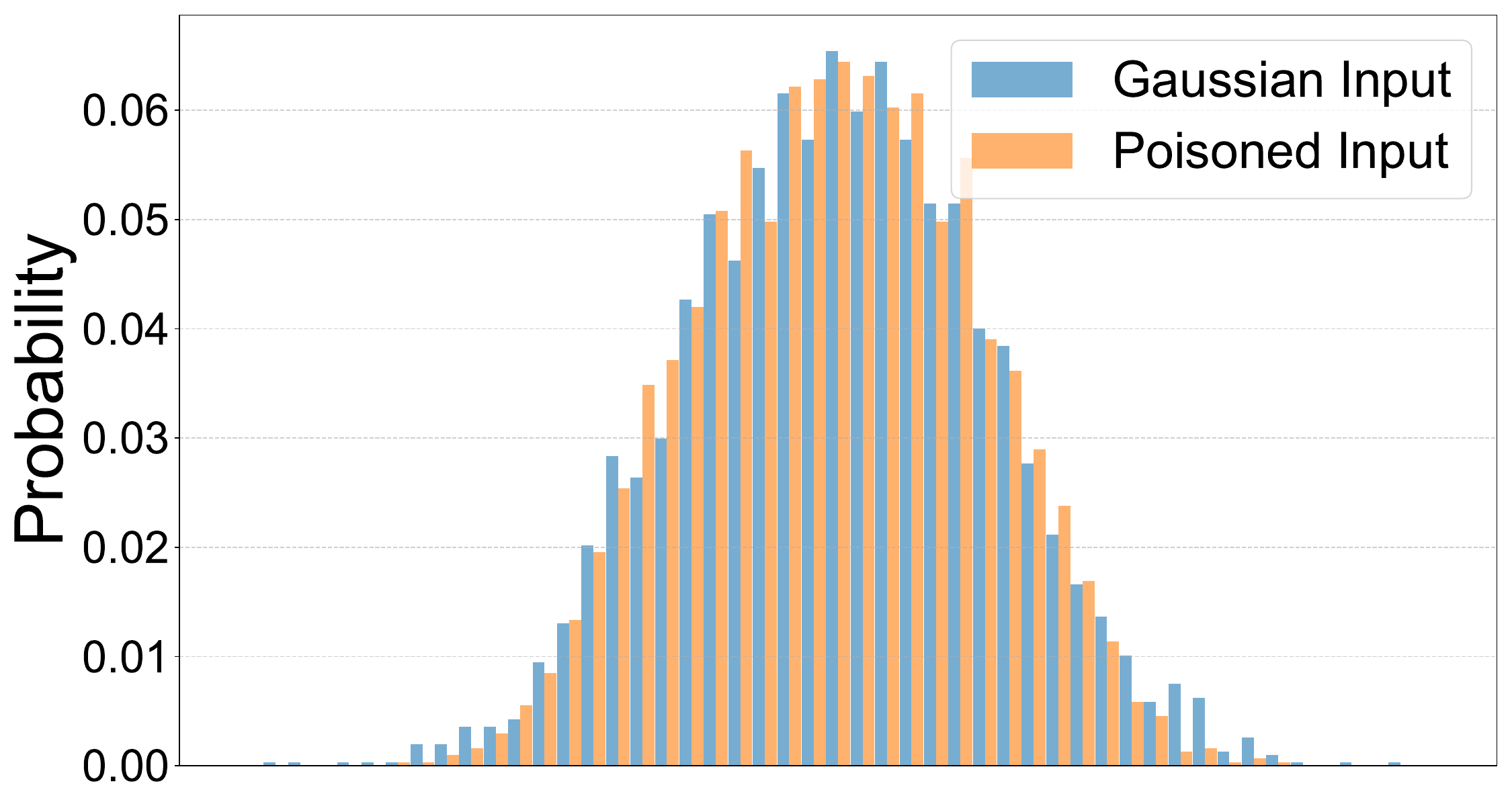}
    \caption{PDD score: 0.023.}
    \label{fig:distribution_ours_b}
  \end{subfigure}
  \caption{Distribution of the poisoned inputs with (a) the fixed trigger \cite{chen2023trojdiff} and (b) our proposed learnable trigger on CIFAR-10 dataset. PDD-oriented optimization brings a much lower PDD score, making our poisoned inputs approach clean Gaussian noise.}
  \label{fig:distribution_ours}
\vspace{-4mm}
\end{figure}

\noindent \textbf{Effect of PDD Optimization.} 
\cref{fig:effect_kl} shows the curve of differentiable PDD score $D_d(\Tilde{\mathbf{x}}_T)$ during the trigger training procedure. It is seen that it steadily decreases as training progresses, and finally this loss reaches below $\phi_{Th}$, indicating that the proposed differentiable histogram $h_d(\cdot)$ is an effective approximation to $h(\cdot)$ when considering the gradient-based optimization. Also, \cref{fig:distribution_ours} illustrates the distribution discrepancy incurred by fixed trigger used in \cite{chen2023trojdiff} and learnable trigger after PDD optimization. It is seen that the proposed PDD-oriented trigger learning brings a much lower PDD score, significantly improving the stealthiness of the backdoor trigger.

\begin{figure}[t]
  \centering
  \begin{subfigure}{0.49\linewidth}
    \includegraphics[width=1\linewidth]{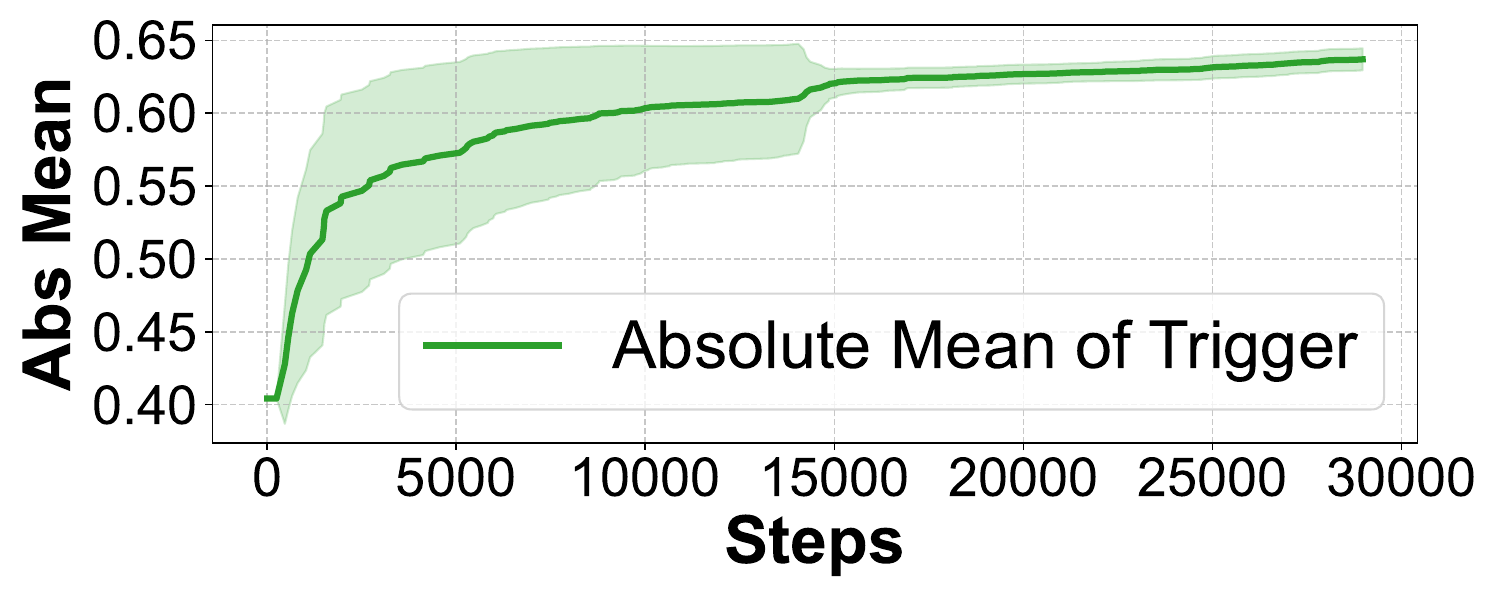}
    \caption{Attack mode: ``Category".}
    \label{fig:effect_kl_absmean_a}
  \end{subfigure}
  \hfill
  \begin{subfigure}{0.49\linewidth}
    \includegraphics[width=1\linewidth]{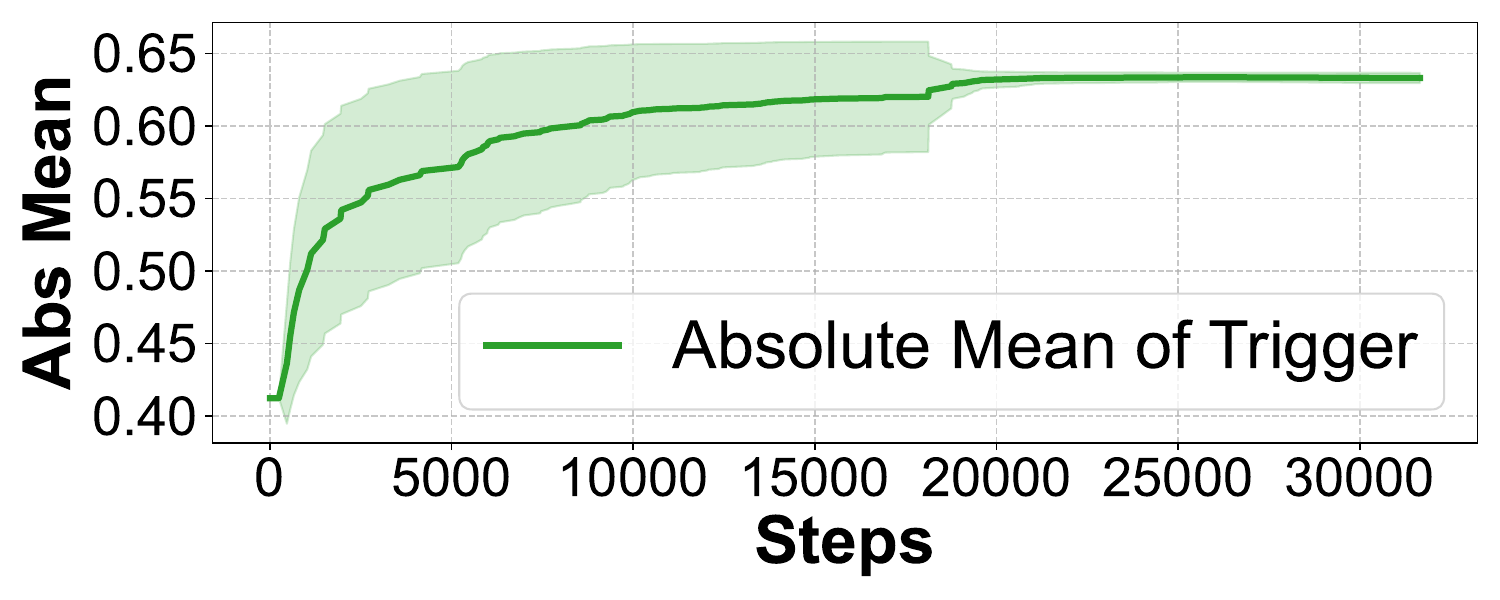}
    \caption{Attack mode: ``Instance".}
    \label{fig:effect_kl_absmean_b}
  \end{subfigure}
  \caption{The curve of absolute mean of trigger values during the PDD-oriented training process.}
  \label{fig:effect_kl_absmean}
\end{figure}

We also observe the variation of the absolute mean of the trigger values when optimizing the PDD scores, as illustrated in \cref{fig:effect_kl_absmean}. Although the mean of trigger values remains around zero, there is a significant variation in absolute mean values. As the training process progresses, the absolute mean increases while the PDD score decreases. This phenomenon leads us to hypothesize that the change in the absolute mean value of the trigger may be related to the process of approaching Gaussian distribution. The modified values seem to mimic the distribution of Gaussian noise. As the PDD scores reach a plateau, the mean absolute value also becomes stable, no longer experiencing significant fluctuations. 


\begin{figure}[t]
  \centering
  \begin{subfigure}{0.49\linewidth}
    \includegraphics[width=1\linewidth]{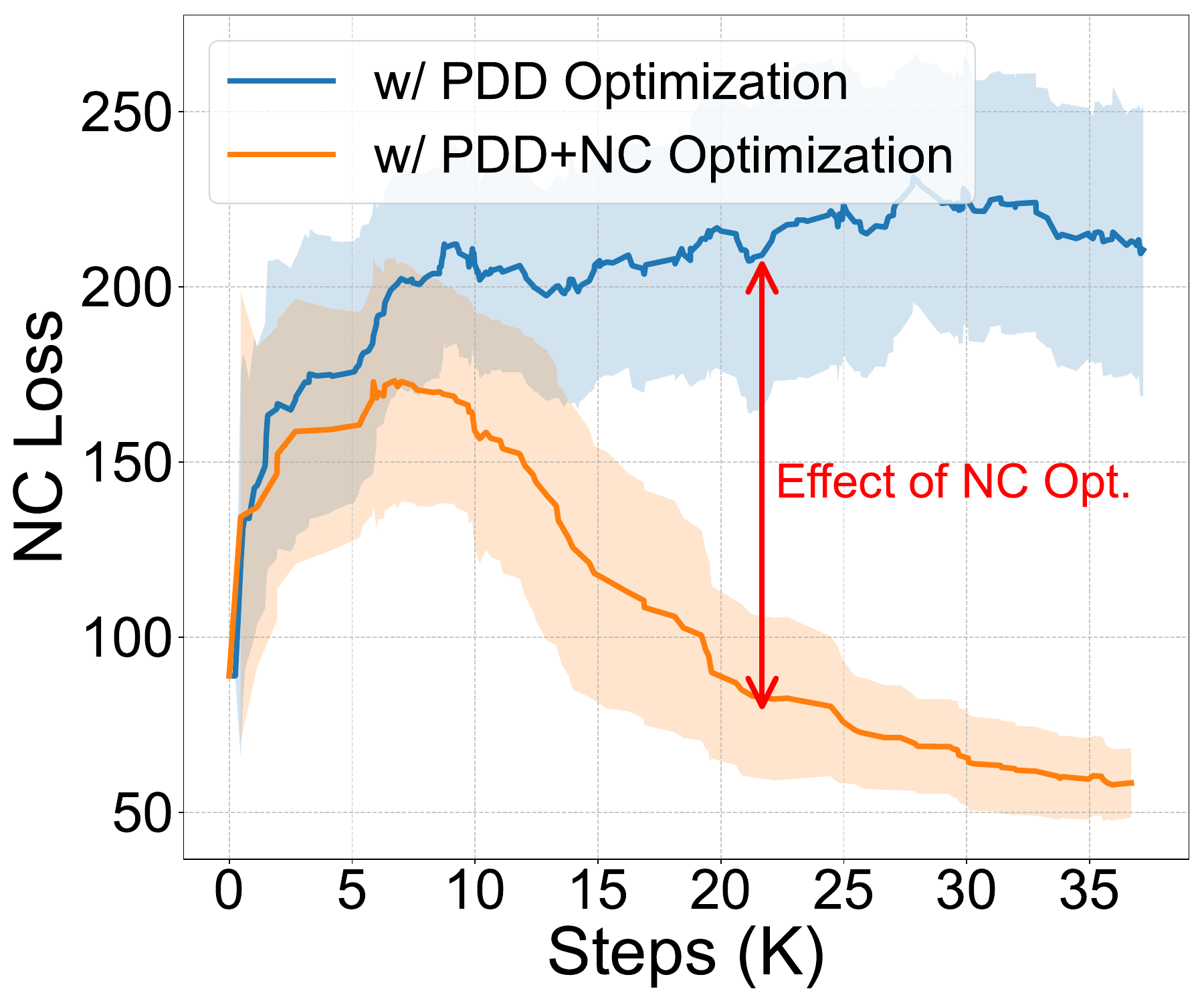}
    \caption{Attack mode: ``Category".}
    \label{fig:effect_mse_nc_a}
  \end{subfigure}
  \hfill
  \begin{subfigure}{0.49\linewidth}
    \includegraphics[width=1\linewidth]{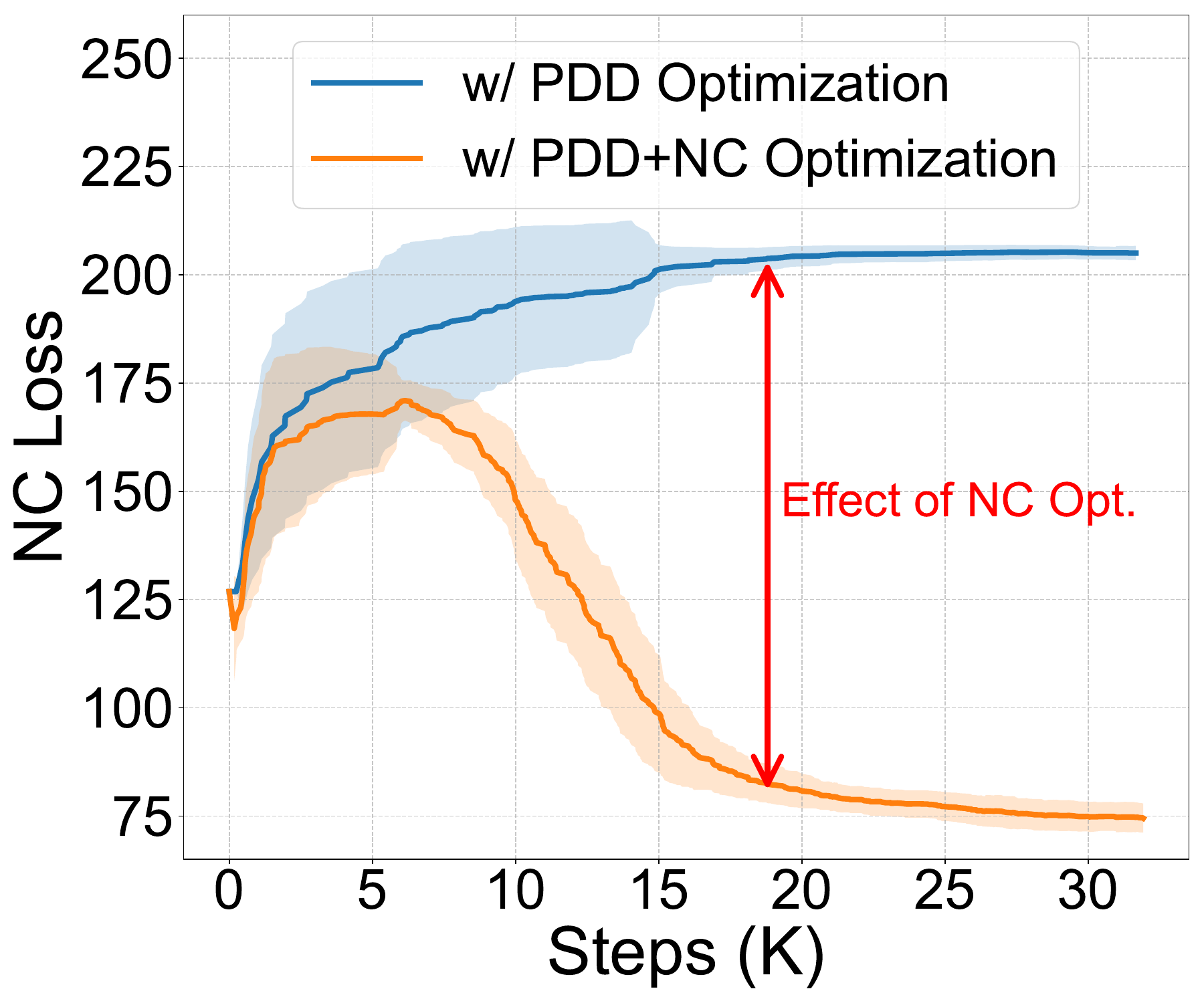}
    \caption{Attack mode: ``Instance".}
    \label{fig:effect_mse_nc_b}
  \end{subfigure}
  \caption{Effect of NC optimization. $\mathcal{L}_{NC}$ in the y-axis denotes the deviation between the predicted and added noises of the diffusion model on CIFAR-10 dataset. Optimizing the trigger $\boldsymbol{\delta}$ with NC loss facilitates the model more easily predicting the noise in the poisoned input noise $\Tilde{\mathbf{x}}_T$. $\mathcal{L}_{NC}$ in attack mode ``instance" is stable since the target image is fixed, causing less variance.}
  \label{fig:effect_mse_nc}
  \vspace{-6mm}
\end{figure}

\begin{figure}[ht]
  \centering
  \begin{subfigure}{0.49\linewidth}
    \includegraphics[width=1\linewidth]{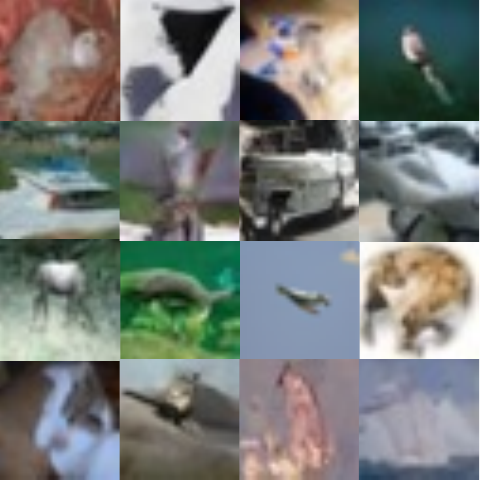}
    \caption{w/ PDD, w/o NC.}
    \label{fig:effect_mse_nc_vis_a}
  \end{subfigure}
  \hfill
  \begin{subfigure}{0.49\linewidth}
    \includegraphics[width=1\linewidth]{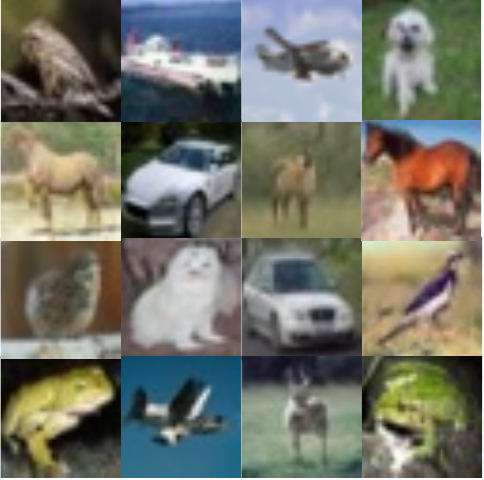}
    \caption{w/ PDD, w/ NC.}
    \label{fig:effect_mse_nc_vis_b}
  \end{subfigure}
  \vfill
    \begin{subfigure}{0.49\linewidth}
    \includegraphics[width=1\linewidth]{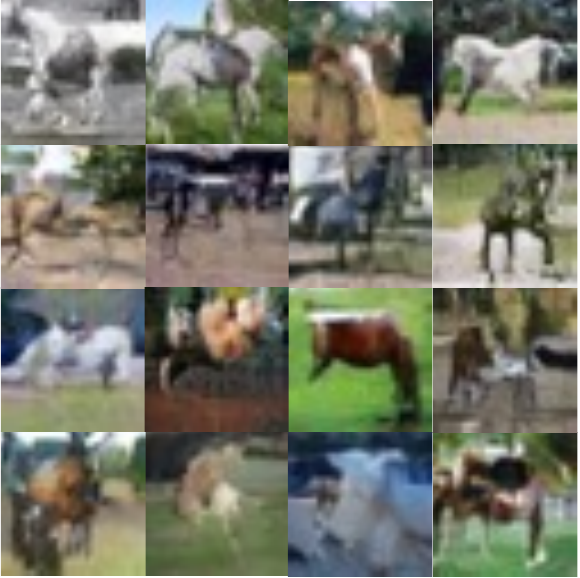}
    \caption{w/ PDD, w/o NC.}
    \label{fig:effect_mse_nc_vis_c}
  \end{subfigure}
  \hfill
  \begin{subfigure}{0.49\linewidth}
    \includegraphics[width=1\linewidth]{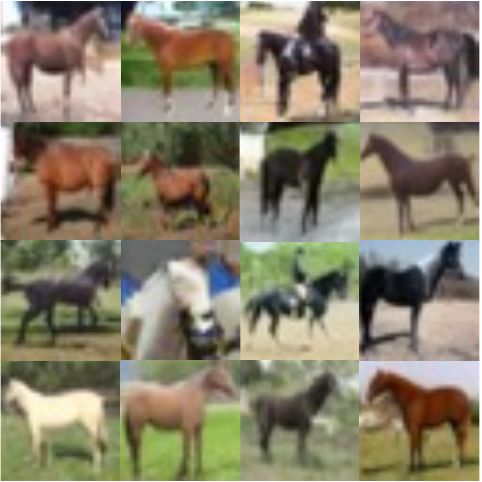}
    \caption{w/ PDD, w/ NC.}
    \label{fig:effect_mse_nc_vis_d}
  \end{subfigure}
  \caption{Visualization of generated CIFAR-10 images with or without NC optimization. Figures (a) and (b) display generated clean images with benign input (i.e., bird, ship, airplane, dog, horse, automobile, deer, frog). Figures (c) and (d) show generated images of the target class (``horse") with poisoned inputs. Figures (b) and (d) represent the images generated from poisoned inputs with the NC-optimized trigger. NC optimization is observed to enhance the quality of image generation in both benign and attack scenarios.}
  \label{fig:effect_mse_nc_vis}
  \vspace{-6mm}
\end{figure}

\noindent \textbf{Effect of NC Optimization.} 
As analyzed in \cref{subsec:NC}, NC loss measures the discrepancy between the added Gaussian noise and the predicted noise at one step. \cref{fig:effect_mse_nc} shows the curve of NC loss as training progresses. It is seen that compared with only emphasizing optimization of PDD, training towards both optimizing both PDD and NC brings a very significant NC loss drop, indicating the strong noise prediction capability and generation of higher-quality images. As shown in \cref{tab:effect_mse_fid_asr}, using NC optimization leads to lower FID scores across different sampling steps. \cref{fig:effect_mse_nc_vis} visualizes the benign and attack performance improvement after considering NC optimization in the trigger learning.


\begin{table}[t]
  \setlength{\tabcolsep}{16pt}
  \centering
  \scalebox{0.8}{
  \begin{tabular}{c|c|ccc}
    \toprule
    & \multirow{2}{*}{Init Loss} & \multicolumn{3}{c}{Sampling Steps, $\eta = 0$} \\
    \cline{3-5} 
    & & 50 & 100 & 200  \\

    \midrule
                     w/o NC & 205.1 & 5.07 & 4.51 & 4.50 \\
    \rowcolor{gray!10} \textbf{w/ NC} & \textbf{74.3} & \textbf{4.99} & \textbf{4.38} & \textbf{4.17} \\
    \rowcolor{gray!10}  \textbf{$\Delta$} & \textbf{130.8} & \textbf{0.08} & \textbf{0.13} & \textbf{0.33} \\
    \bottomrule
  \end{tabular}
  }
  \caption{FID results of backdoored diffusion model with the trigger optimized with or without NC optimization. The results are evaluated with different sampling steps on the CIFAR-10 dataset.}
  \label{tab:effect_mse_fid_asr}
  \vspace{-4mm}
\end{table}

\section{Conclusion}

This paper performs systematic studies on the detectability of the Trojan input for the backdoored diffusion model, from both defender and attacker sides. It proposes a distribution discrepancy-based trigger detection mechanism as well as the corresponding detection-evading trigger design solution. Evaluation results show the high protection and attack performance of our proposed approaches.

\clearpage
{
    \small
    \bibliographystyle{ieeenat_fullname}
    \bibliography{main}
}


\clearpage
\setcounter{page}{1}
\maketitlesupplementary





\section{Additional Ablation Studies}

\subsection{Effect of Various Smoothness Factors}

\begin{figure}[ht]
  \centering
  
  \begin{subfigure}{0.49\linewidth}
    \includegraphics[width=1\linewidth]{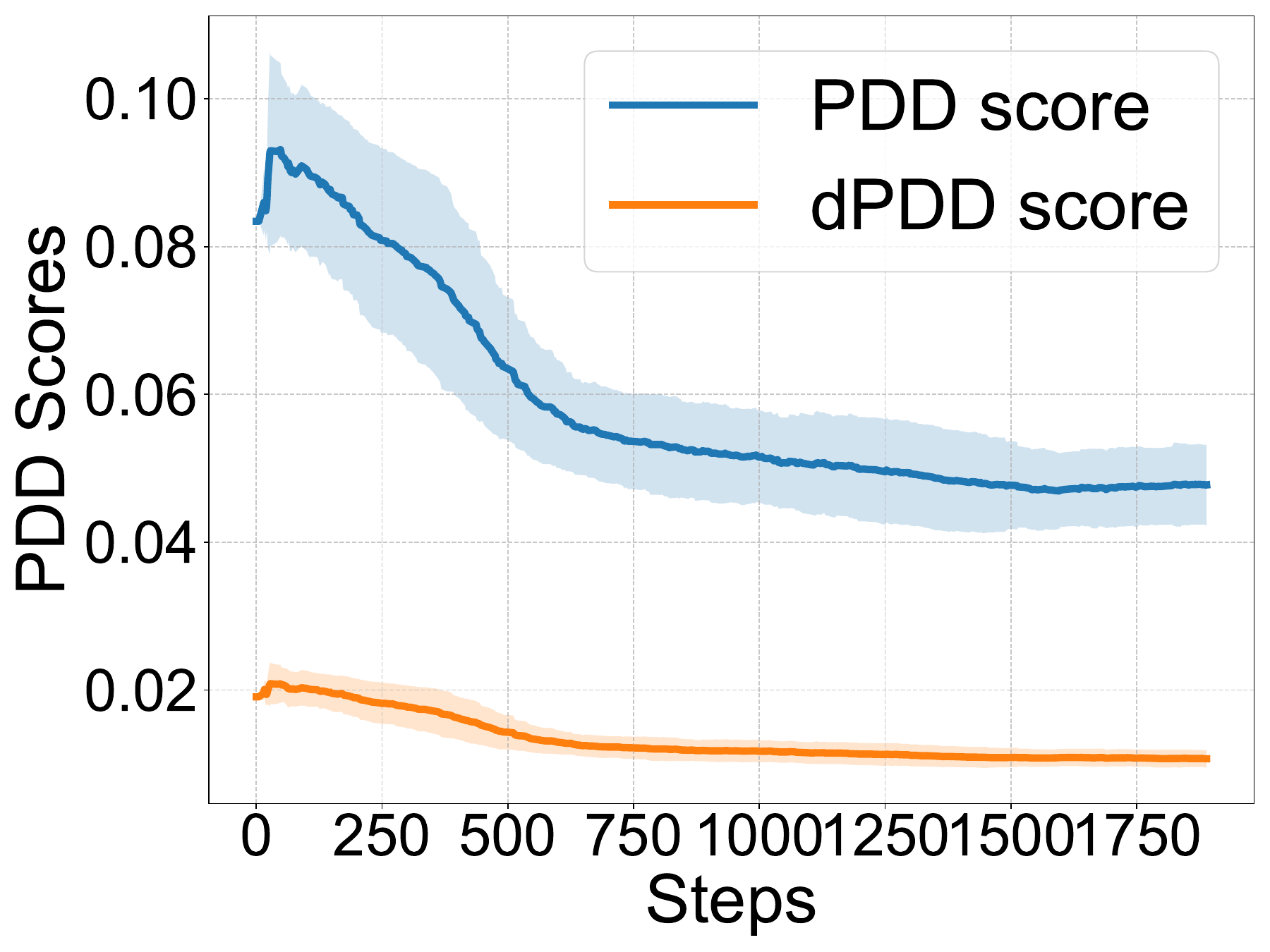}
    \caption{$\omega=2$.}
    \label{fig:omega_2}
  \end{subfigure}
  \hfill
  \begin{subfigure}{0.49\linewidth}
    \includegraphics[width=1\linewidth]{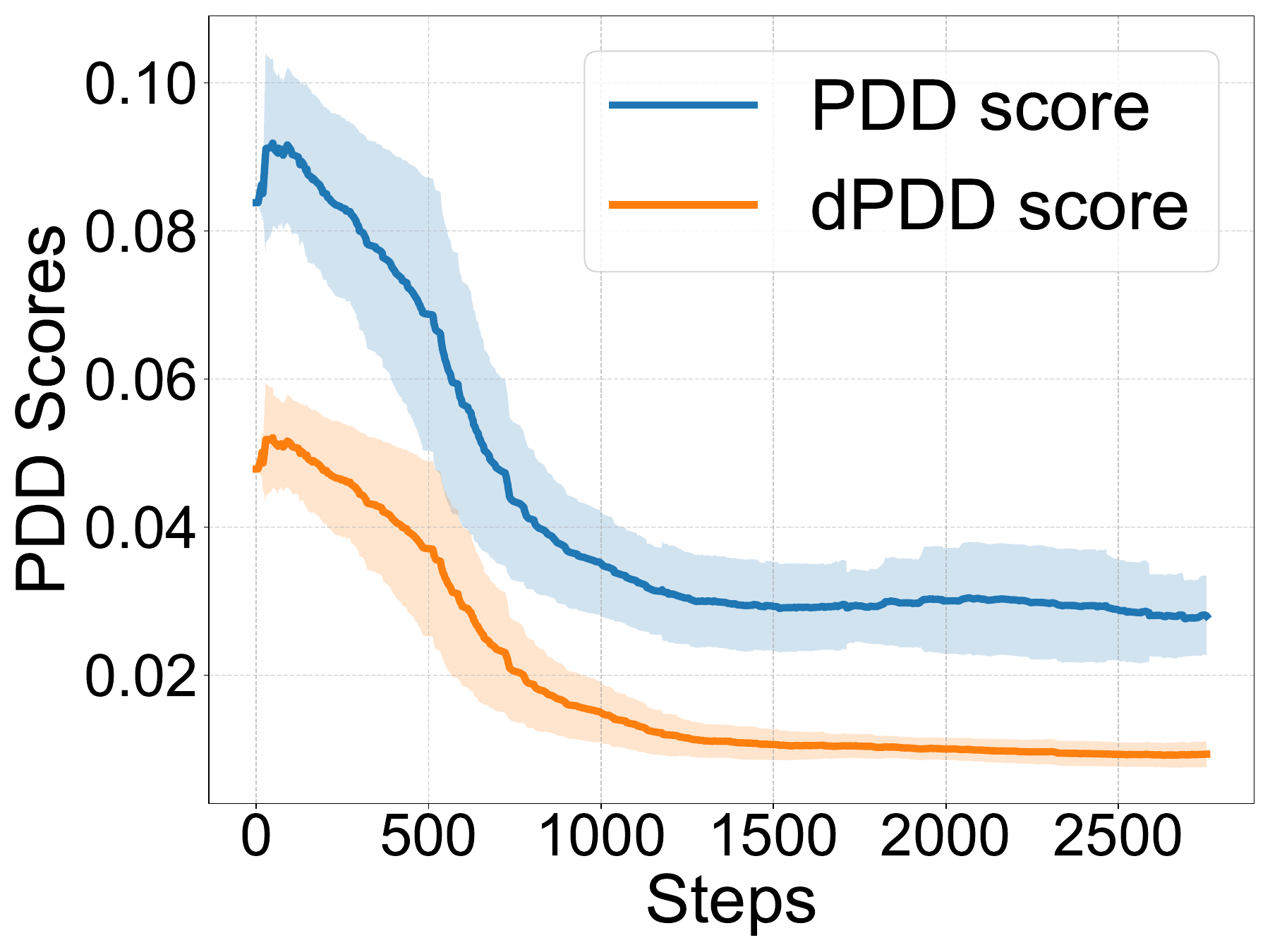}
    \caption{$\omega=4$.}
    \label{fig:omega_4}
  \end{subfigure}
  
  \vfill
  
  \begin{subfigure}{0.49\linewidth}
    \includegraphics[width=1\linewidth]{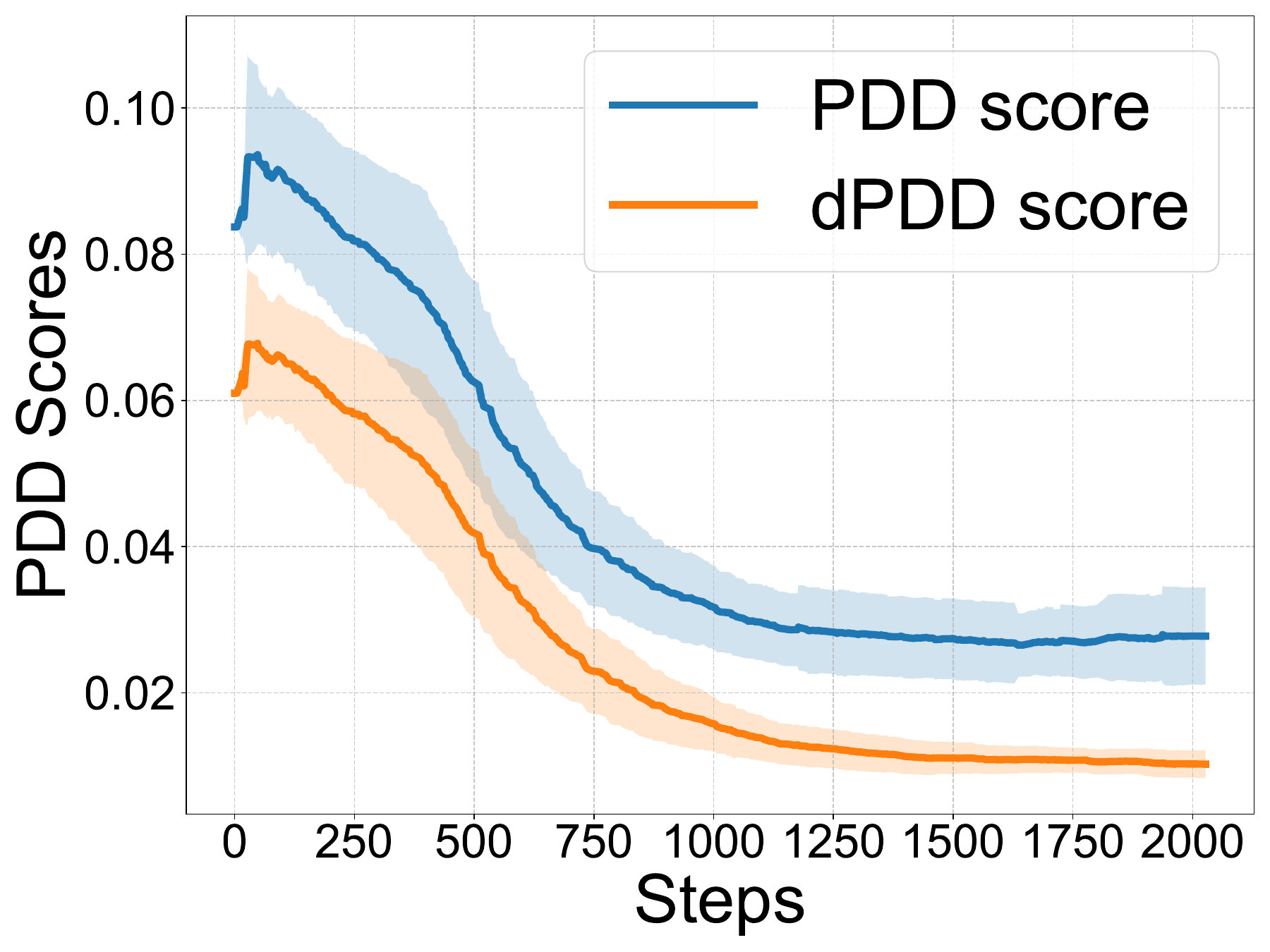}
    \caption{$\omega=6$.}
    \label{fig:omega_6}
  \end{subfigure}
  \hfill
  \begin{subfigure}{0.49\linewidth}
    \includegraphics[width=1\linewidth]{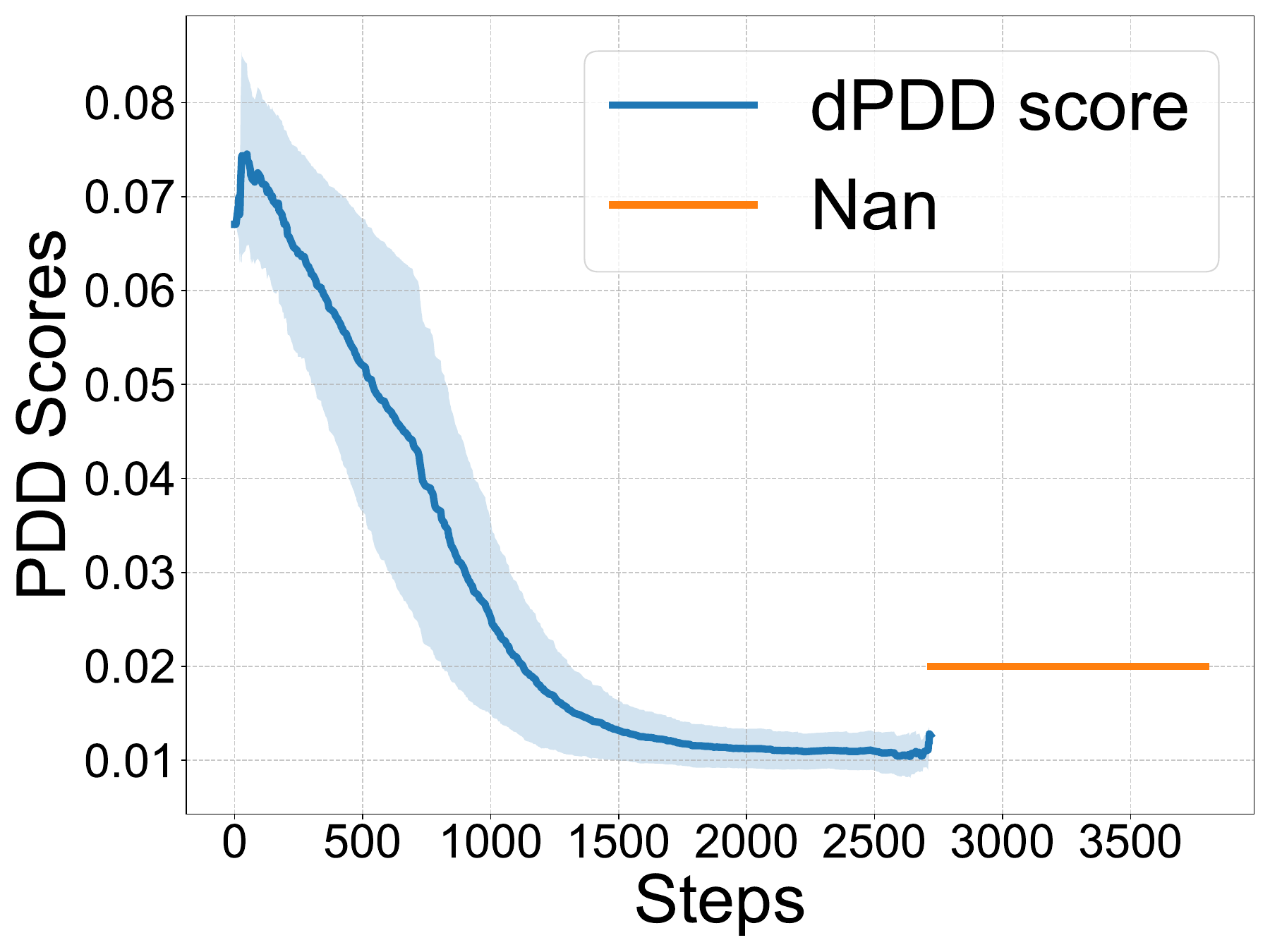}
    \caption{$\omega=8$.}
    \label{fig:omega_8}
  \end{subfigure}
  
  \caption{Curve of PDD score $D(\Tilde{\mathbf{x}}_T)$ and differentiable PDD score $D_d(\Tilde{\mathbf{x}}_T)$ when the trigger is trained with the PDD loss $\mathcal{L}_{dPDD}$. Effect of $\omega$, the smoothness factor of the differential histogram function.}
  \label{fig:omega}
\end{figure}

To explore the effect of the smoothness factor $\omega$ of the differential histogram function $h_d(\cdot)$, we conduct several PDD optimization experiments using various $\omega$. \cref{fig:omega} shows the curve of the PDD score $D(\Tilde{\mathbf{x}}_T)$ and differentiable PDD score $D_d(\Tilde{\mathbf{x}}_T)$ during the trigger PDD optimization procedure. 

When $\omega$ is too small, such as $\omega=2$ in \cref{fig:omega_2}, the approximation of differentiable PDD score $D_d(\Tilde{\mathbf{x}}_T)$ becomes less accurate. This inaccuracy results in a very small differential PDD score, even at the initialization step. Then, it leads to a larger gap between the regular PDD score $D(\Tilde{\mathbf{x}}_T)$ and differentiable PDD score $D_d(\Tilde{\mathbf{x}}_T)$. This larger gap is undesirable for optimizing the desired distribution pass rate and ASR.


When $\omega$ is too large (e.g., $\omega=8$), the differential histogram function exhibits non-smooth characteristics. This can be seen in \cref{fig:omega_8}, where after 2.5K steps, the PDD scores become undefined (denoted as ``NaN"). This phenomenon is directly attributed to the non-smooth nature of the differential histogram function with larger $\omega$ values. Consequently, a larger $\omega$ not only hinders but also halts the optimization process of the desired PDD score.

In our experiments, we choose $\omega = 6$ to strike a balance between minimizing the gaps in PDD score and maintaining the feasibility of optimization.





\subsection{Sampling Steps and Training Epochs}

To investigate how training and sampling steps affect the performance of a stealthy backdoor diffusion model, we conduct training on CIFAR-10 and CelebA datasets using the ``instance" attack mode. The models are trained with varying training steps (50k, 75k, 100k) and sampling steps (10, 20, 50, 100, 200). According to the results presented in  \cref{tab:sampling_steps}, an increase in training epochs generally maintains or slightly decreases the FID while simultaneously enhancing attack performance. Regarding sampling steps, we observe that larger number of sampling steps leads to improved performance in both benign and attack scenarios. This aligns with the property of a standard diffusion model, where increasing sampling steps tend to yield better overall performance.

\begin{table}[ht]
  \setlength{\tabcolsep}{5pt}
  \centering
  \scalebox{0.8}{
  \begin{tabular}{c|c|ccccc}
    \toprule
     Training & \multirow{2}{*}{Metric} & \multicolumn{5}{c}{Sampling Steps} \\
     \hhline{|~|~|-|-|-|-|-|}
     Epochs & & 10 & 20 & 50 & 100 & 200 \\
    \hhline{|-|-|-|-|-|-|-|}
    \multicolumn{7}{c}{CIFAR-10} \\
    \hline
    \multirow{2}{*}{50k} & FID & 14.74 & 7.42 & 5.00 & 4.39 & 4.17 \\
                         & MSE & 1.53e-4 & 1.03e-4 & 8.37e-5 & 8.10e-5 & 7.34e-5 \\
    \hline
    \multirow{2}{*}{75k} & FID & 14.37 & 7.33 & 4.98 & 4.34 & 4.17 \\
                         & MSE & 1.07e-4 & 7.12e-5 & 5.69e-5 & 5.52e-5 & 5.02e-5 \\
    \hline
    \multirow{2}{*}{100k} & FID & 14.51 & 7.38 & 4.99 & 4.38 & 4.17 \\
                         & MSE & 8.20e-5 & 5.43e-5 & 4.32e-5 & 4.19e-5 & 3.80e-5 \\
    \midrule
    \multicolumn{7}{c}{CelebA} \\
    \hline
    \multirow{2}{*}{50k} & FID & 2.85e-3 & 2.02e-3 & 1.71e-3 & 1.70e-3 & 1.59e-3 \\
                         & MSE & 13.31 & 8.23 & 6.35 & 5.85 & 5.75 \\
    \hline
    \multirow{2}{*}{75k} & FID & 1.43e-3 & 1.04e-3 & 9.01e-4 & 8.98e-4 & 8.55e-4 \\
                         & MSE & 13.24 & 8.17 & 6.28 & 5.84 & 5.71 \\
    \hline
    \multirow{2}{*}{100k} & FID & 1.08e-3 & 7.98e-4 & 7.01e-4 & 6.99e-4 & 6.69e-4 \\
                         & MSE & 13.25 & 8.14 & 6.23 & 5.82 & 5.72 \\
    \bottomrule
  \end{tabular}
  }
  \caption{Results of different sampling steps on various training epochs on CIFAR-10 and CelebA datasets with attack mode ``instance". For the benign performance, FID reflects the quality of benign images. For the attack performance, the ASR equals that of the main results, and MSE is measured to reflect the subtle change exactly according to different sampling steps and training epochs. }
  \label{tab:sampling_steps}
\end{table}

\section{Additional Visualization Results}
\label{sec:additional_vis_results}

\subsection{CIFAR-10 Dataset}

\begin{figure}[ht]
    \centering
    \includegraphics[width=1\linewidth]{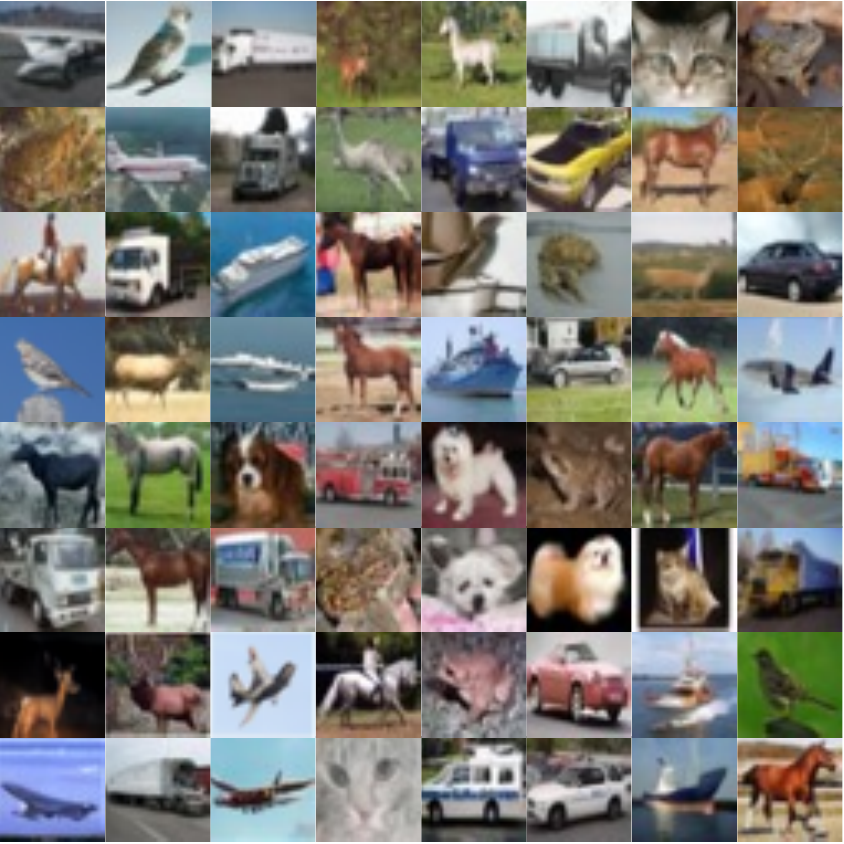}
    \caption{Generated benign images in Attack Mode ``Category" with DDPM model on CIFAR-10 dataset.}
    \label{fig:appendix_vis_ddpm_1}
\end{figure}

\begin{figure}[t]
    \centering
    \includegraphics[width=1\linewidth]{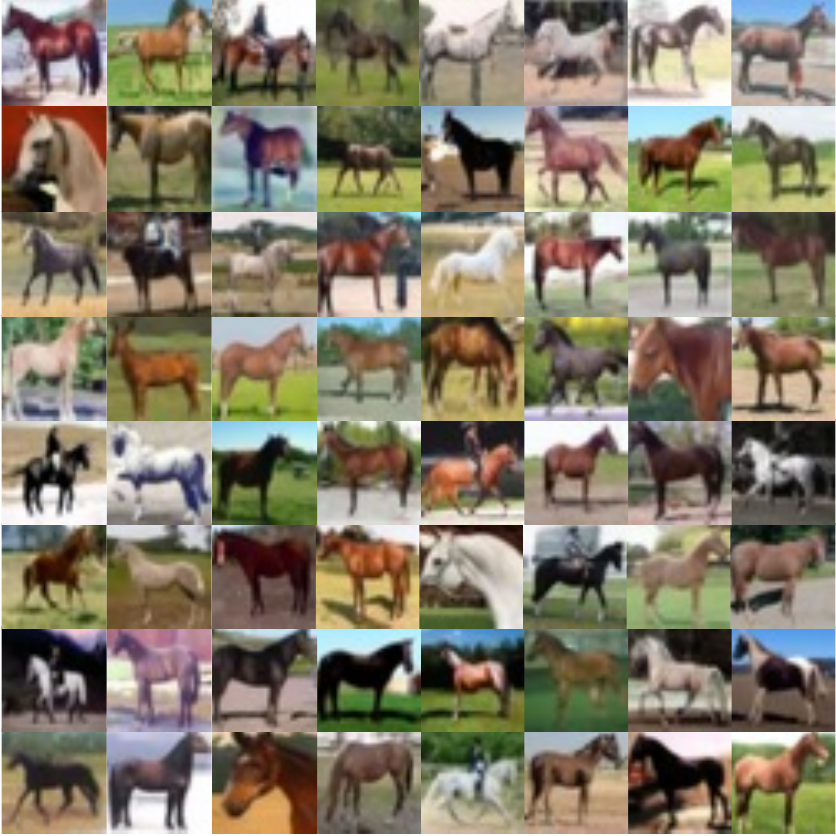}
    \caption{Generated target images in Attack Mode ``Category" with DDPM model on CIFAR-10 dataset. The target category is the horse.}
    \label{fig:appendix_vis_ddpm_2}
\end{figure}

\begin{figure}[t]
    \centering
    \includegraphics[width=1\linewidth]{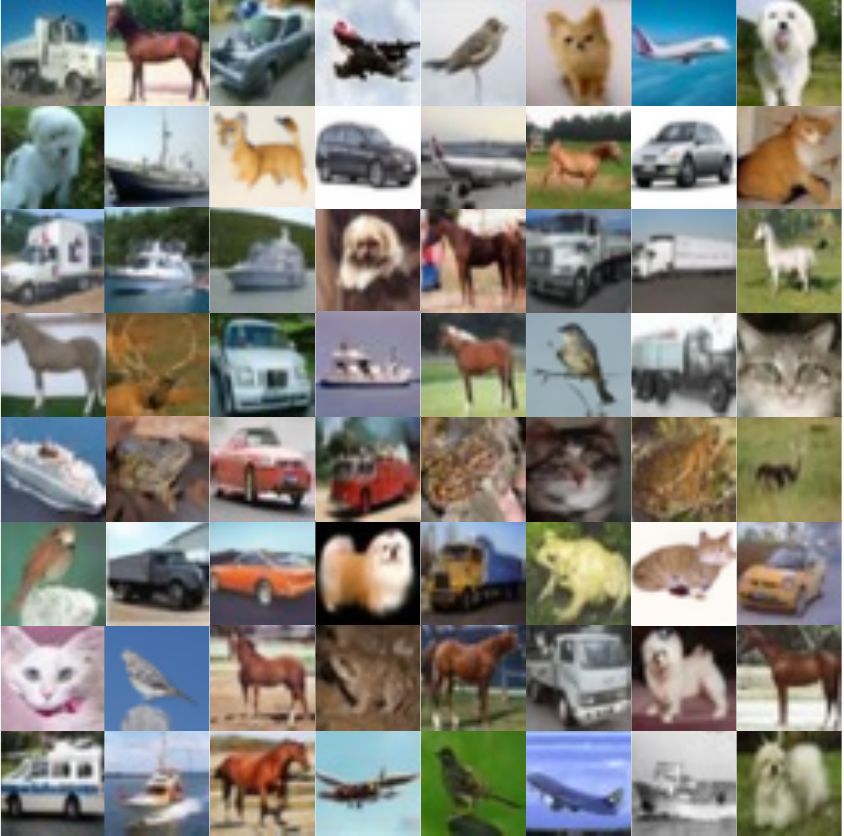}
    \caption{Generated benign images in Attack Mode ``Instance" with DDPM model on CIFAR-10 dataset.}
    \label{fig:appendix_vis_ddpm_3}
\end{figure}

\begin{figure}[t]
    \centering
    \includegraphics[width=1\linewidth]{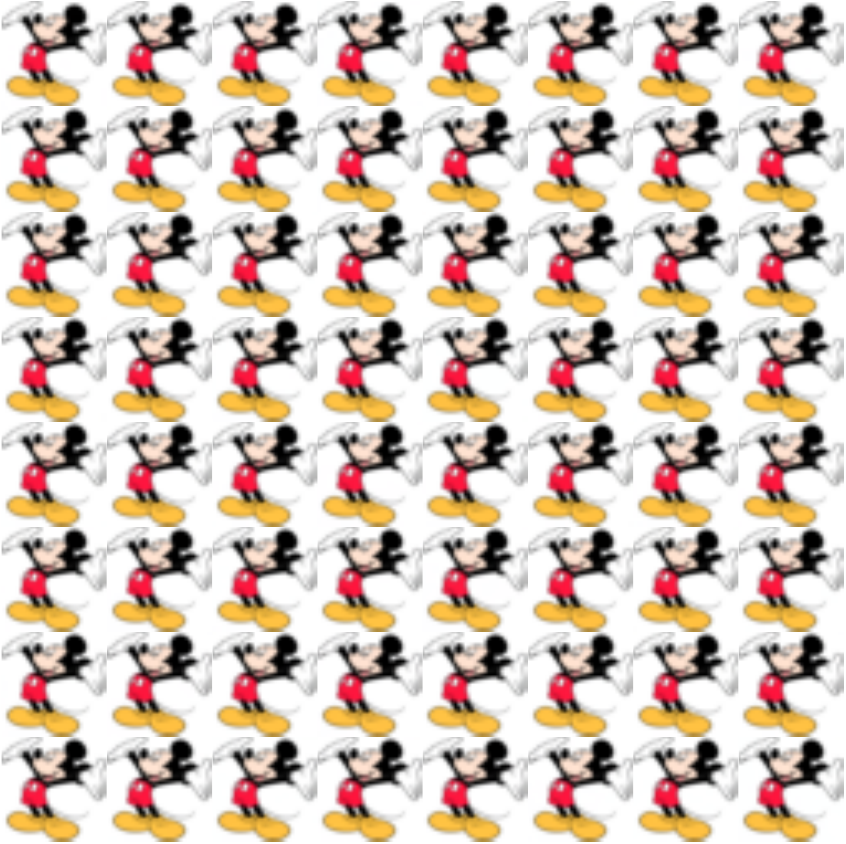}
    \caption{Generated target images in Attack Mode ``Instance" with DDPM model on CIFAR-10 dataset. The target instance is the Michy Mouse image.}
    \label{fig:appendix_vis_ddpm_4}
\end{figure}

\noindent \textbf{DDPM.} 
\cref{fig:appendix_vis_ddpm_1,fig:appendix_vis_ddpm_2,fig:appendix_vis_ddpm_3,fig:appendix_vis_ddpm_4} illustrate some of the generated images from our backdoored DDPM model with benign and poisoned noise inputs on CIFAR-10 dataset. For the attack mode ``Category", the target category is the horse. For the attack mode ``Instance", the target instance is a Michy Mouse image. 

\begin{figure}[t]
    \centering
    \includegraphics[width=1\linewidth]{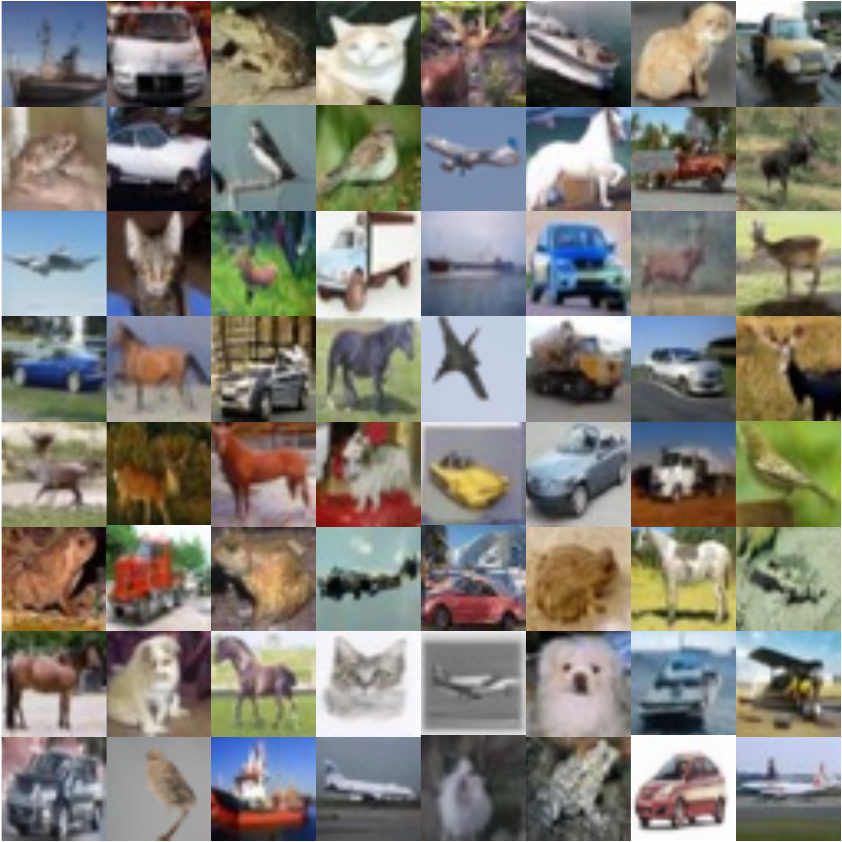}
    \caption{Generated benign images in Attack Mode ``Category" with DDIM model on CIFAR-10 dataset.}
    \label{fig:appendix_vis_ddim_1}
\end{figure}

\begin{figure}[t]
    \centering
    \includegraphics[width=1\linewidth]{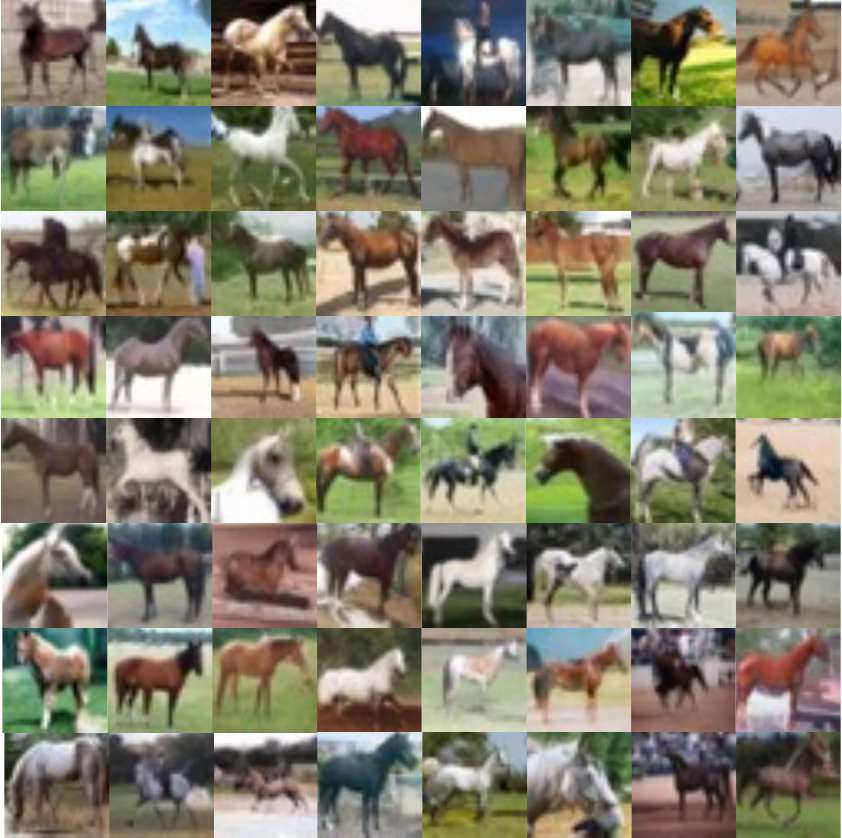}
    \caption{Generated target images in Attack Mode ``Category" with DDIM model on CIFAR-10 dataset. The target category is the horse.}
    \label{fig:appendix_vis_ddim_2}
\end{figure}

\begin{figure}[t]
    \centering
    \includegraphics[width=1\linewidth]{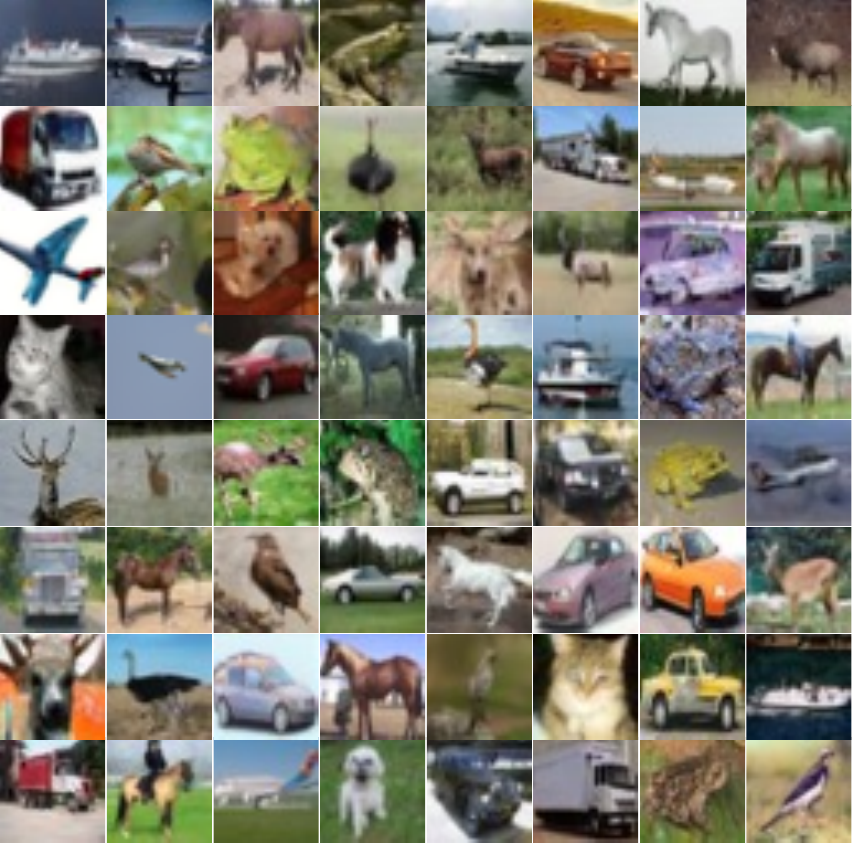}
    \caption{Generated benign images in Attack Mode ``Instance" with DDIM model on CIFAR-10 dataset.}
    \label{fig:appendix_vis_ddim_3}
\end{figure}

\begin{figure}[t]
    \centering
    \includegraphics[width=1\linewidth]{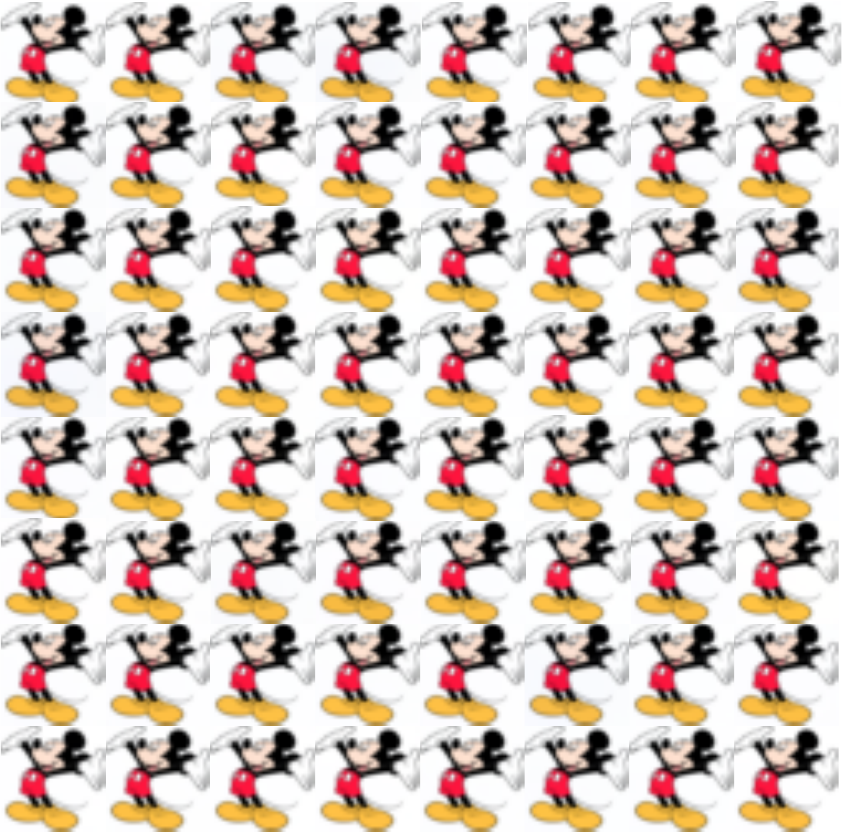}
    \caption{Generated target images in Attack Mode ``Instance" with DDIM model on CIFAR-10 dataset. The target instance is the Michy Mouse image.}
    \label{fig:appendix_vis_ddim_4}
\end{figure}

\noindent \textbf{DDIM.} \cref{fig:appendix_vis_ddim_1,fig:appendix_vis_ddim_2,fig:appendix_vis_ddim_3,fig:appendix_vis_ddim_4} illustrate some of the generated images from our backdoored DDIM model with benign and poisoned noise inputs on CIFAR-10 dataset. For the attack mode ``Category", the target category is the horse. For the attack mode ``Instance", the target instance is a Michy Mouse image. 

\subsection{CelebA Dataset}

\begin{figure}[t]
    \centering
    \includegraphics[width=1\linewidth]{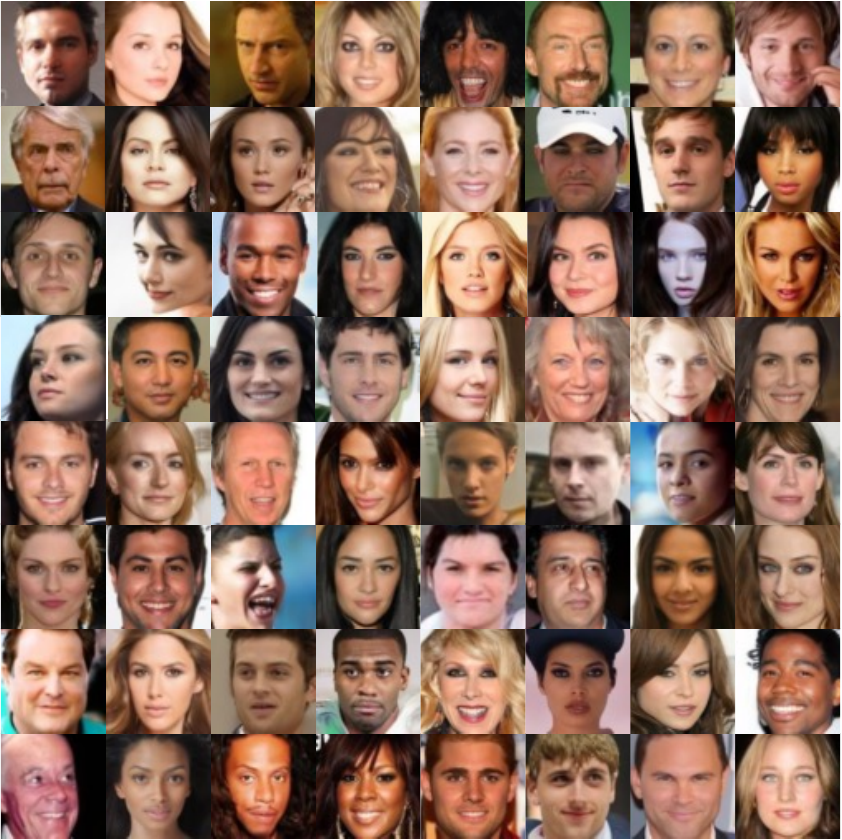}
    \caption{Generated benign images in Attack Mode ``Category" with DDPM model on CelebA dataset.}
    \label{fig:appendix_vis_ddpm_5}
\end{figure}

\begin{figure}[t]
    \centering
    \includegraphics[width=1\linewidth]{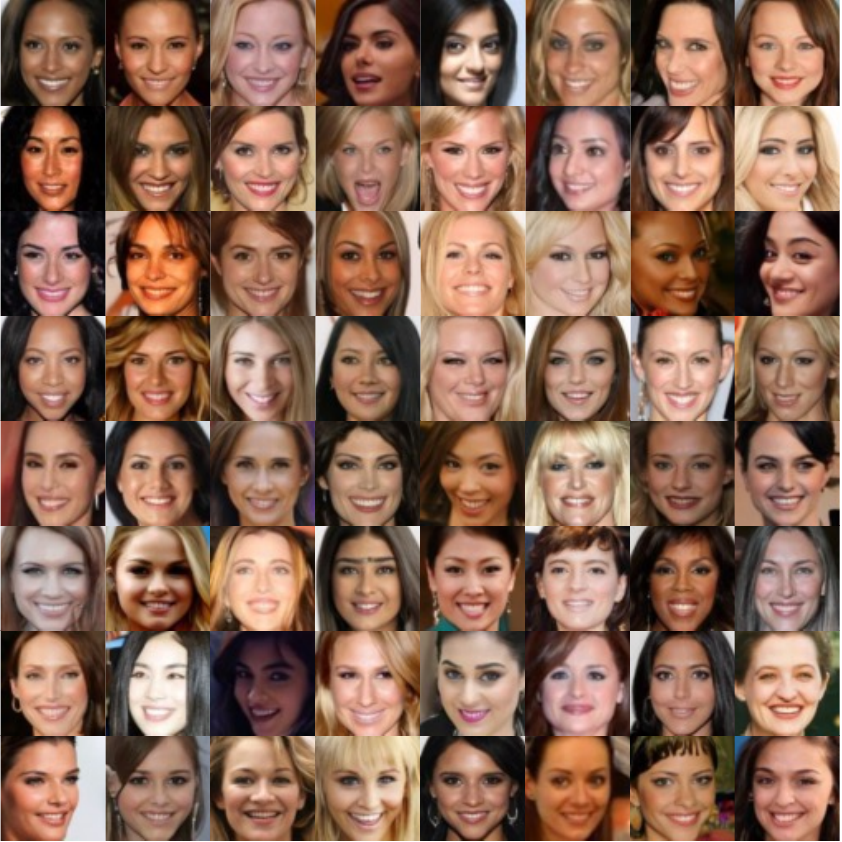}
    \caption{Generated target images in Attack Mode ``Category" with DDPM model on CelebA dataset. The target category is the faces with heavy makeup, mouth slightly open and smiling.}
    \label{fig:appendix_vis_ddpm_6}
\end{figure}

\begin{figure}[t]
    \centering
    \includegraphics[width=1\linewidth]{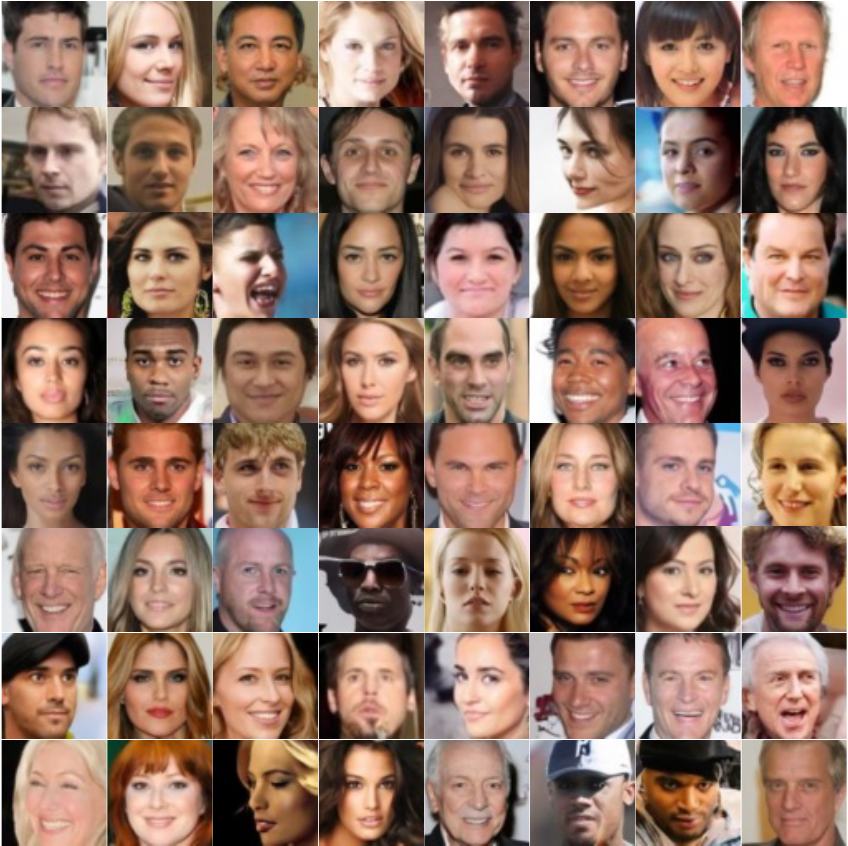}
    \caption{Generated benign images in Attack Mode ``Instance" with DDPM model on CelebA dataset.}
    \label{fig:appendix_vis_ddpm_7}
\end{figure}

\begin{figure}[t]
    \centering
    \includegraphics[width=1\linewidth]{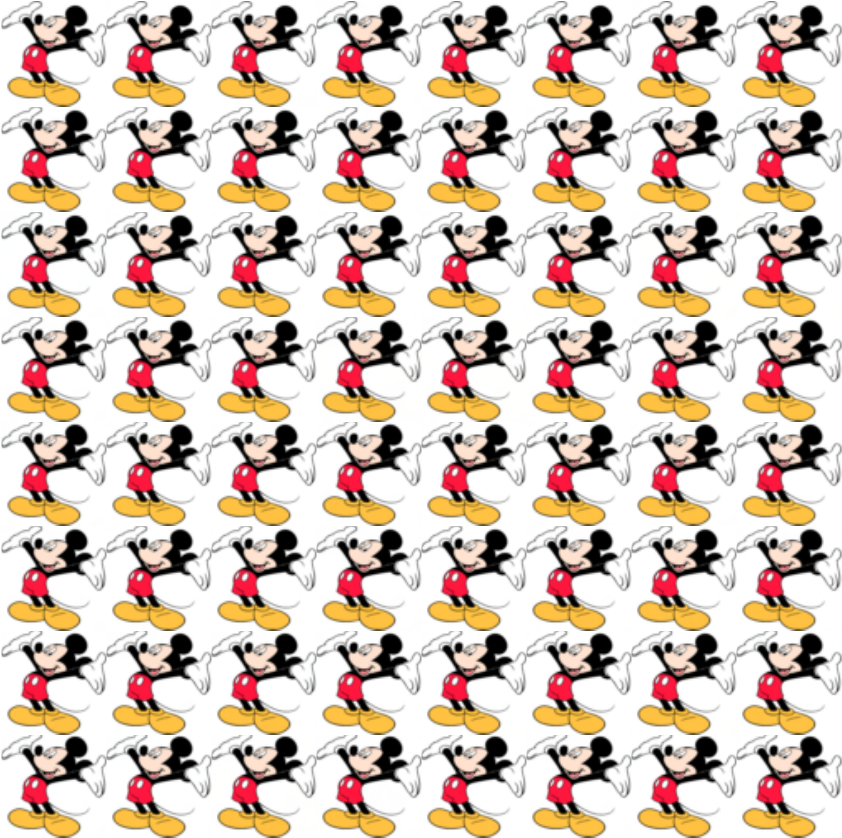}
    \caption{Generated target images in Attack Mode ``Instance" with DDPM model on CelebA dataset. The target instance is the Michy Mouse image.}
    \label{fig:appendix_vis_ddpm_8}
\end{figure}

\noindent \textbf{DDPM.} \cref{fig:appendix_vis_ddpm_5,fig:appendix_vis_ddpm_6,fig:appendix_vis_ddpm_7,fig:appendix_vis_ddpm_8} illustrate some of the generated images from our backdoored DDPM model with benign and poisoned noise inputs on CelebA dataset. For the attack mode ``Category", the target category is the faces with heavy makeup, mouth slightly open and smiling. For the attack mode ``Instance", the target instance is a Michy Mouse image. 

\begin{figure}[t]
    \centering
    \includegraphics[width=1\linewidth]{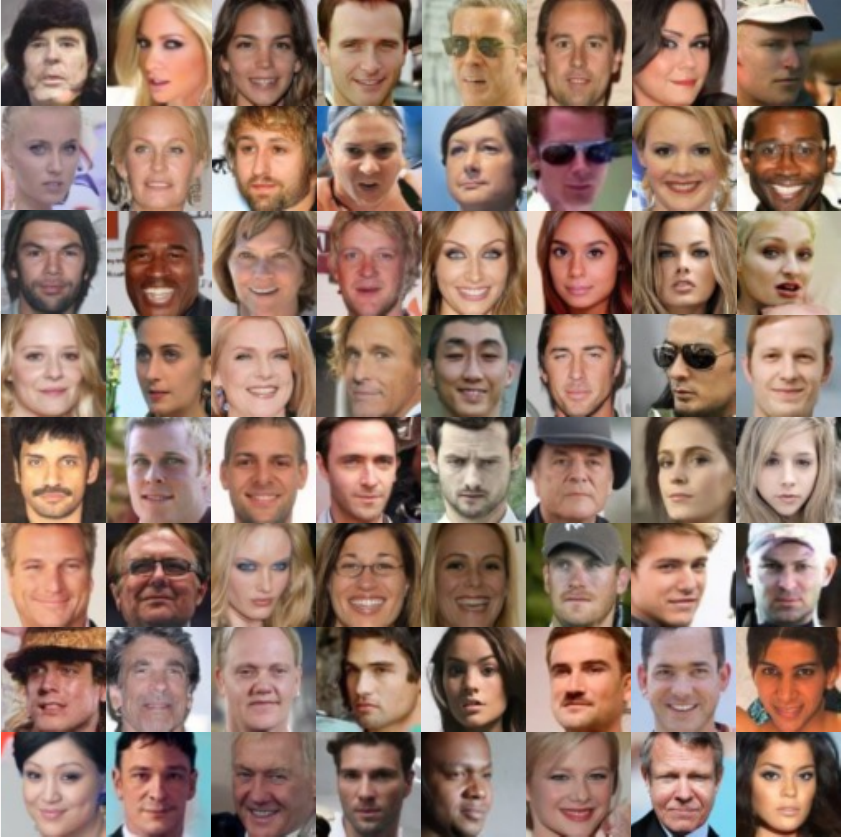}
    \caption{Generated benign images in Attack Mode ``Category" with DDIM model on CelebA dataset.}
    \label{fig:appendix_vis_ddim_5}
\end{figure}

\begin{figure}[t]
    \centering
    \includegraphics[width=1\linewidth]{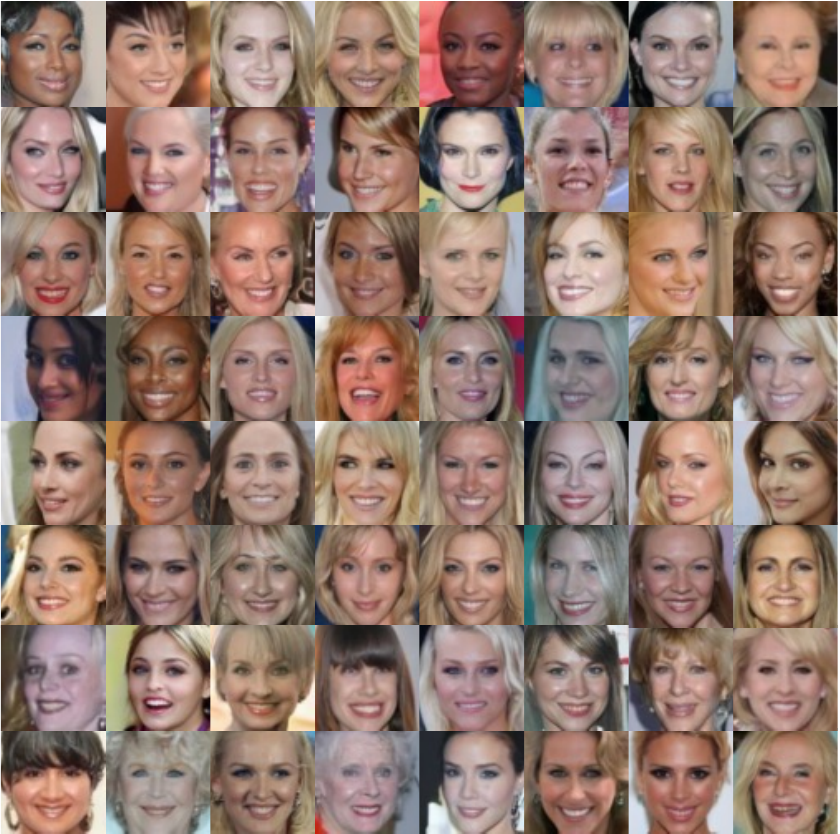}
    \caption{Generated target images in Attack Mode ``Category" with DDIM model on CelebA dataset. The target category is the faces with heavy makeup, mouth slightly open and smiling.}
    \label{fig:appendix_vis_ddim_6}
\end{figure}

\begin{figure}[t]
    \centering
    \includegraphics[width=1\linewidth]{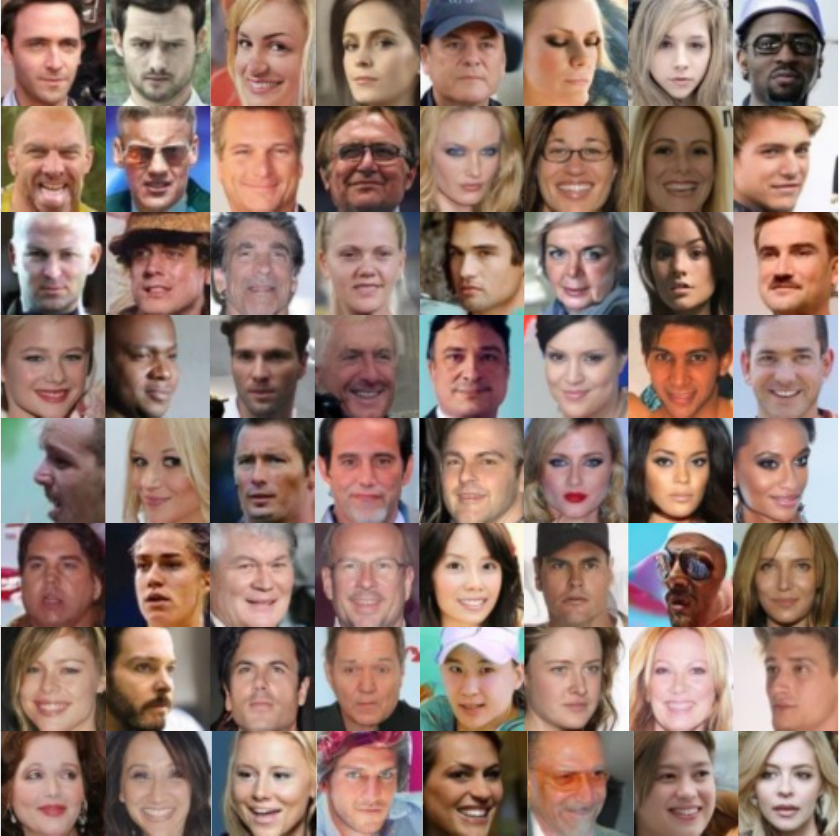}
    \caption{Generated benign images in Attack Mode ``Instance" with DDIM model on CelebA dataset.}
    \label{fig:appendix_vis_ddim_7}
\end{figure}

\begin{figure}[t]
    \centering
    \includegraphics[width=1\linewidth]{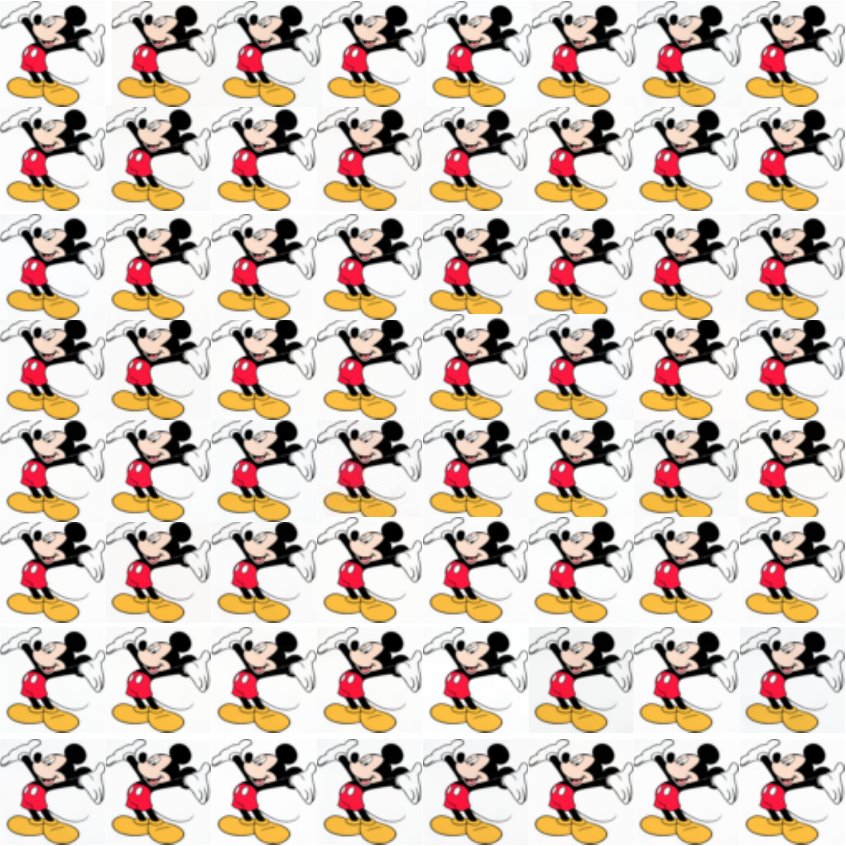}
    \caption{Generated target images in Attack Mode ``Instance" with DDIM model on CelebA dataset. The target instance is the Michy Mouse image.}
    \label{fig:appendix_vis_ddim_8}
\end{figure}

\noindent \textbf{DDIM.} \cref{fig:appendix_vis_ddim_5,fig:appendix_vis_ddim_6,fig:appendix_vis_ddim_7,fig:appendix_vis_ddim_8} illustrate some of the generated images from our backdoored DDIM model with benign and poisoned noise inputs on CelebA dataset. For the attack mode ``Category", the target category is the faces with heavy makeup, mouth slightly open and smiling. For the attack mode ``Instance", the target instance is a Michy Mouse image. 


\end{document}